\documentclass[conference]{IEEEtran}
\IEEEoverridecommandlockouts
\usepackage{optidef}
\usepackage{authblk}
\usepackage{cite}
\usepackage{booktabs}
\usepackage{amsmath,amssymb,amsfonts}
\usepackage[hang,flushmargin]{footmisc}
\usepackage{multirow}
\usepackage{float}
\usepackage{textcomp}
\usepackage{lscape}
\usepackage{url}
\usepackage{enumerate}
\usepackage{caption}
\usepackage{subcaption}
\usepackage{algpseudocode}
\usepackage{graphicx}
\usepackage{textcomp}
\usepackage{xcolor}
\usepackage{array}
\usepackage{mathtools,algorithm}
\usepackage{tabularx}
\usepackage{cuted}
\usepackage{enumitem}
\usepackage{authblk}

 \newcommand{\comm}[1]{}  

\newcommand{\CASE}[1]{\STATE \textbf{case} #1\textbf{:} \begin{ALC@g}}
\newcommand{\ENDCASE}{\end{ALC@g}}

\newcommand{\DEFAULT}{\STATE \textbf{default:} \begin{ALC@g}}
\newcommand{\ENDDEFAULT}{\end{ALC@g}}
\newcommand{\DEFAULTLINE}[1]{\STATE \textbf{default:} }
\newcolumntype{L}[1]{>{\raggedright\let\newline\\\arraybackslash\hspace{0pt}}m{#1}}
\newcolumntype{C}[1]{>{\centering\let\newline\\\arraybackslash\hspace{0pt}}m{#1}}
\newcolumntype{R}[1]{>{\raggedleft\let\newline\\\arraybackslash\hspace{0pt}}m{#1}}
\def\BibTeX{{\rm B\kern-.05em{\sc i\kern-.025em b}\kern-.08em
    T\kern-.1667em\lower.7ex\hbox{E}\kern-.125emX}}

\begin{document}

\title{Optical Networks}

\author[1,2,*]{Varsha Lohani}
\author[1,2]{Anjali Sharma}
\author[1,2]{Yatindra Nath Singh }
\author[1]{\\Kumari Akansha}
\author[1]{Baljinder Singh Heera}
\author[1,2]{Pallavi Athe}

\affil[1]{Department of Electrical Engineering, Indian Institute of Technology Kanpur, Kanpur, India}
\affil[2]{YRRNA Systems Lab}

\affil[*]{lohani.varsha7@gmail.com}

%

\maketitle
\begin{abstract}
Optical networks play a crucial role in today's digital topography, enabling the high-speed and reliable transmission of vast amounts of data over optical fibre for long distances. This paper provides an overview of optical networks, especially emphasising on their evolution with time. 
\end{abstract}

\section{Introduction}
Communication has always played a vital role in the advancement of society. It allowed the exchange of ideas without any limit on space and time. Communication in forms such as verbal (face-to-face, audio, telephone, video calls etc.), non-verbal (facial expressions, eye contact etc.), or written (books, newspapers, messages etc.) helped establish the information flow paths within society \cite{evolution}. Communication modes have evolved due to technological improvements in the last few decades. This section will explain how communication changed over time and how optical fibre networks contributed to the development of modern-day communication Networks.
    
The optical telegraph in 1794 by the Chappe Brothers, the independent inventions of Cooke-Wheatstone electric telegraphs and the electric telegraph by Samuel Morse in 1837 \cite{telegraphy} were some of the initial breakthroughs in the field of telecommunication. The electric telegraph was used to transmit messages over long distances through electrical pulses using codes formed by binary symbols. After telegraphy, another significant milestone in telecommunication was the invention of the telephone by Alexander Graham Bell in 1876. A telephone transforms voice into electronic signals, which are transferred via electrical wires over long distances. It was further developed commercially for local and long-distance calls. In order to share costly long-distance lines to reduce the cost per unit of voice call, manual telephone exchanges evolved. In 1891, the stepping switch invented by Almon Brown Strowger contributed to telephone switch automation, leading to more traffic-handling capabilities and fewer errors in circuit setup \cite{strowger}. These networks further evolved through the use of crossbar switches to improve reliability. In these networks, voice traffic was transmitted using copper cables.

Alexander Graham Bell also patented an optical telephone in 1880, also known as Photophone. However, Photophone never happened. In the 1960s, Abraham Van Heel made a notable contribution to optical fibre design. He covered a bare glass fibre with a transparent coating such that the light was confined within the fiber and did not leak out. The covering is known as cladding, and its refractive index is normally less than that of the fibre core. However, these were only used for short distances because glass-clad fibres had an attenuation of about 1 decibel (dB) per meter. In the same decade, Charles Kuen Kao developed optical fibre for use in the field of telecommunication \cite{fb1}. Earlier, attenuation was the biggest challenge in the optical transmission medium. Kao removed impurities from the glass and made it highly transparent. The attenuation was reduced significantly in particular wavelength windows. Improving optical fibre communication over the years involved optimizing bandwidth, data rate, signal transmission quality, dispersion compensation, noise reduction, and various other technical aspects to enable faster, more reliable, and more efficient communication through optical fibres.

In recent years, Internet traffic has expanded exponentially due to the proliferation of a range of technologies, including multimedia and cloud computing applications. Cisco Annual Internet Survey report had predicted that nearly two-thirds of the world’s population will have Internet access by 2023, and the peak fixed world broadband speeds will reach 110.4 Mbps in 2023, up from 45.9 Mbps in 2018 \cite{cisco}. ITU-D reports verified that as of the year 2023, 67$\%$ of total world population is online \cite{ITU-D}.  Such large traffic volumes are feasible; thanks to optical networks, which are evolving further. Continuous improvements and innovations have led to better performance of network endpoint devices. It has further forced improvements in optical communication network capacities to meet the high bandwidth demands.
    
Optical fibers are generally made of plastic or glass. Key benefits of using optical fiber are high capacity, lower bit error rates, tolerances to noise and electromagnetic interference, large bandwidth-distance product, and difficulty in signal tapping \cite{fb2}. Almost all wired communication networks nowadays use optical fibre, and are commonly known as \textbf{Optical Networks} \cite{rama}. In an optical network, though transmission is optical, the switching at nodes can be optical, electronic, or hybrid. Further, we can have circuit, packet or burst switching in an optical network. An optical network is classified as an All-Optical Network (AON) if the transmitted signal remains optical from source to destination without any intermediate Optical-Electrical-Optical (O-E-O) conversion. 
    
Optical networks promise various advantages over conventional communication networks. They can provide long-distance communication with Terabits per second capacity while sharing infrastructure among various network nodes.\comm{\cite{rama}} Though light signals travel with the speed of light in glass medium\footnote{$\dfrac{c}{n} = 2 \times 10^{8} m/s$ for $n = 1.5$}, the actual data transfer rate is governed by modulation bandwidth.
    
The electromagnetic spectrum in the range from 850 nm to 1675 nm is available for communication in glass fibres \cite{fb2} for different applications e.g., 850nm-band is used for multimode, O-band between 1260-1360. Opting the entire optical fibre bandwidth for setting up only a single lightpath in optical networks results in its inefficient usage. The interface electronics limit the single lightpath capacity to much less than the available bandwidth in fibre. Optical networks have evolved based on how the transmission, switching, and multiplexing techniques have changed over time to use this available bandwidth more efficiently. 
    
Conventionally in optical networks, the optical domain was used only for transmission and capacity provisioning, while the switching and other network functions at nodes were managed in the electronic domain. Examples of such optical networks are SONET/SDH networks. Over time, they have evolved to more sophisticated systems to meet exponentially growing traffic needs \cite{book1}.
    
Over the years, scientists and network engineers realized that optical networks are much more capable than just providing point-to-point communication, which led to the development of next-generation optical networks. These networks provide routing and switching at the optical layer and are designed to operate with centralized or distributed algorithms. 

Due to the limitation of electrical-to-optical and optical-to-electrical interfaces, bandwidth in optical networks needs to be harnessed using Multiplexing Techniques. Among the possible techniques, Wavelength Division Multiplexing (WDM) is simplest and hence most commonly used to partition the optical spectrum of a few THz\footnote{In this paper we are considering 4THz of optical bandwidth.} into 50 GHz, or 100 GHz \cite{itut} channels. WDM creates parallel channels, each of which can carry one connection, thus providing more capacity compared to single-channel networks. These networks are also called Fixed-Grid Optical Networks\footnote{WDM-based Optical Networks or Wavelength Switched Optical Networks}. Due to the fixed wavelength grid, the connection requests above a certain bandwidth cannot be accommodated. Also, connection requests with lower bandwidth lead to wastage of optical bandwidth as the whole wavelength channel needs to be allocated irrespective of the requested bandwidth. 
    
Flexible-grid Optical Networks\footnote{Elastic Optical Networks or SLICE (Slice-able Optical Networks)} provide more efficient use of the optical spectrum compared to the fixed-grid networks \cite{EON1}. They use Optical Orthogonal Frequency Division Multiplexing to handle the variable bandwidth connection requests effectively and efficiently. This multiplexing technique helps to create flexible channels of sizes which are integer multiples of 12.5GHz \cite{itut}. Super channels are created by combining many contiguous channels, thereby accommodating connection requests of varying sizes. 
    
WDM-based Optical Networks and Elastic Optical Networks implementation also pose challenges such as Routing and Resources Assignment (RRA)\footnote{depending upon the type of network}, Resource Conversion, Network Control and Management, and Survivability against failures, to name a few \cite{RWA11}.

\subsection{Single Mode Optical Fiber for Core Optical Networks}
In the simplest model, the light can only be launched at certain discrete angles in an optical fibre. This means only finite modes are allowed inside an optical fiber, and the light launched at any other angle is lost and not guided. 

The multiple discrete modes have different propagation constants and hence different groups velocities as they traverse inside an optical fibre, because of their different allowable launching angles. The different group velocities lead to pulse broadening\footnote{also known as Dispersion.} as different copies of same signal reach the receiver at slightly different time instants. Single-mode fibre is preferred to avoid pulse broadening due to multiple modes (modal dispersion). In Single Mode Optical Fiber, the diameter of the core is kept around $\sim$ 5-10 $\mu$m so that only a single mode for the wavelength of interest can traverse through it.

There is another way of reducing the pulse broadening, i.e., by gradual variation of the refractive index of the core. Such types of fibers are known as graded index multi-mode fibers. However, it does not completely mitigate the impact of pulse broadening. Therefore, we opt only for Single Mode Optical Fibers for long-haul communication.
    
\subsection{Paper Organization}
The paper has been organized as follows. Section II discusses various types of optical networks, from fixed to flexible grids. The routing and resource assignment, and survivability techniques in these networks are discussed in sections III and IV, respectively. Section V reviews the work done in the literature to improve the optical networks' resource utilization. The ways to improve the capacity of the optical networks such that they can support higher data rates are discussed in section VI. Section VII provides various prospective areas that need to be investigated to further enhance the performance of the optical networks.

\section{Types of Optical Networks}
\subsection{Fixed-grid: Wavelength Division Multiplexing based Optical Networks}
\begin{figure}[H]
    \begin{center}
        \includegraphics[width=\linewidth]{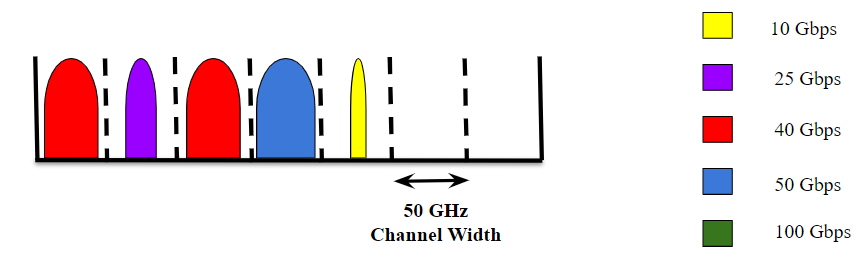}
        \caption{50 GHz Fixed Grid Optical Networks for mixed line rate services.}
        \label{fig:WDM2}
    \end{center}
\end{figure}

An Optical Network can use Wavelength Division Multiplexing (WDM) to segment the spectrum in optical fibre links into multiple parallel channels (50 GHz or 100 GHz channel width). The circuits setup using wavelengths from a source to a destination are also called lightpaths. These are set up by assigning WDM channels for transmission, as shown in Figure \ref{fig:WDM2}. According to ITU-T G.694.1, for a channel width of 50 GHz in an optical fibre, the allowed frequencies in THz are \cite{itut}
\begin{equation}
193.1 + n*0.05, 
\end{equation} 
where n is an integer.

\begin{figure}[H]
    \begin{center}
        \includegraphics[width=\linewidth]{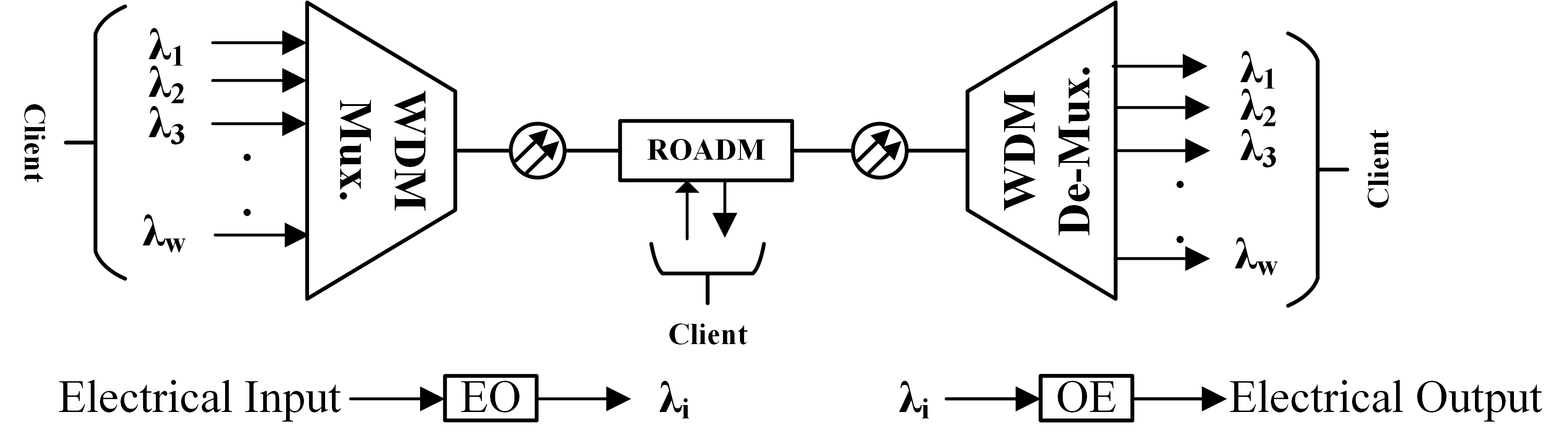}
        \caption{Wavelength Division Multiplexing based transmissions with Multiplexer, Reconfigurable Optical Add Drop Multiplexer (ROADM), De-Multiplexer, \textit{W} Wavelengths ($\lambda_{1}, \lambda_{2},.., \lambda_{w}$), Electrical to Optical Converters (EO), and   Optical to Electrical Converters (OE) at the two ends.}
        \label{fig:WDM1}
    \end{center}
\end{figure}

A wavelength, $\lambda$, corresponds to a carrier frequency $\nu$ ($\nu = \dfrac{c}{n \times \lambda})$ on which signal is modulated. Here $c$ is the speed of light ($3 \times 10^8 m/s$) in vacuum and $n$ is the effective index of glass fiber. One transmitter is needed to generate the signal on one wavelength. Therefore, \textit{W} transmitters are needed for \textit{W} wavelengths on the input side. The signals from the transmitters are multiplexed together and passed through a fibre link. At the output, the WDM signals are de-multiplexed into \textit{W} individual signals and guided to the desired receivers. At the receiver, the conversion of the optical signal into an electronic domain occurs, as shown in Figure \ref{fig:WDM1}.

Customers can communicate with each other over end-to-end optical (WDM) channels called lightpaths in a WDM fibre-based network. A lightpath will pass through multiple fibre links to provide a circuit-switched path between two distant nodes. An optical bypass is given to each lightpath at each intermediate node. Reconfigurable Optical Add-Drop Multiplexer (ROADM) and Optical Cross-Connects (OXCs) are core components of any WDM network that either bypass the wavelength circuits or terminate onto an OLT (Optical Line Terminal) unit. The ROADMs are low-loss, low-cost devices that add/drop multiple desired signals to/from the network. On the other hand, Optical Cross-Connects (OXCs) consist of an $N \times N$ optical switch with \textit{N} input and \textit{N} output fibres. Such an optical switch makes the network All-Optical, i.e., the signals within the system remain optical without any conversion to the electronic domain. The OXCs are responsible for switching the optical signals on one wavelength in an input fibre to same wavelength on an outgoing fibre \cite{book1}. 

A wavelength-based Optical Network functions as a Circuit Switched Network. For data transmission from Source Node to Destination Node, first, a connection is set up, and resources are reserved for the duration of the connection. The connection is released after the connection holding time is over. Here, the resource is the wavelength, which must be same from source to destination. This is the Wavelength Continuity Constraint. The problem of setting up maximum-paths for a given demand matrix with wavelength continuity constraint is called Routing and Wavelength Assignment (RWA). It has been elaborated in the next section. 

The WDM optical network's main drawback is its fixed wavelength grid, which cannot support lightpath requests larger than the channel size.  We need flexible grid networks that can work beyond the fixed channel size limits to accommodate high-capacity demand. In the next sub-section, we review these next-generation optical networks, i.e., the flexible grid networks. These are also called Elastic Optical Networks (EON).

\subsection{Flexible-grid: Elastic Optical Networks}
Elastic Optical Networks (EON)\footnote{a.k.a. Flexible grid Optical Networks or Sliceable Optical Networks} can easily fulfil increasing bandwidth demands using flexible data rates, based on flexible spectrum allocation. They also have low signal attenuation, low signal distortion, and low power requirement (as some sub-carriers can be switched off when not required), less usage of material, less space requirement and higher spectral efficiency.

\begin{figure}[H]
    \begin{center}
        \includegraphics[width=\linewidth]{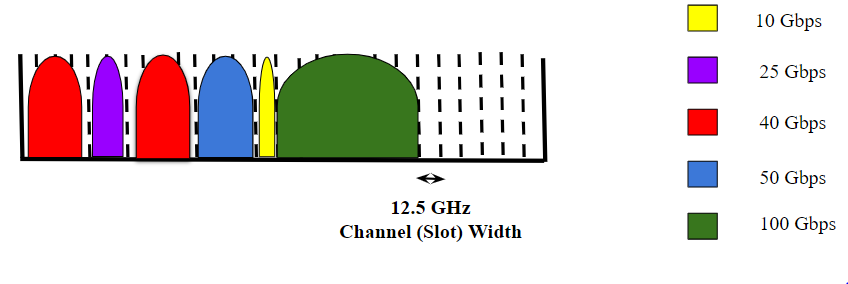}
        \caption{12.5 GHz Grid for Elastic Optical Networks for mixed line rate services.}
        \label{fig:EON1}
    \end{center}
\end{figure}

M. Jinno \textit{et. al,} in\cite{EON1}, proposed flexibility in the optical fibre link spectrum by changing the multiplexing technique from Dense Wavelength Division Multiplexing (WDM)  to Optical-Orthogonal Frequency Division Multiplexing (O-OFDM). As shown in Figure \ref{fig:EON1}, the optical channel size is further reduced to 12.5 GHz with O-OFDM, improving the granularity of bandwidth allocation. O-OFDM-based EON is very efficient due to its flexible combining of basic spectral slots to form superchannels support.

For a flexible grid, the allowed frequency slots  range in THz is defined by ITU-T, where the central frequency of a slot is given by

\begin{equation}
193.1 + n*0.00625,
\end{equation}
where n is an integer. The allocated slot width in GHz in elastic optical networks can be
\begin{equation}
\text{12.5 GHz} * m,
\end{equation} 
where m is a positive integer. The slots allocated to different requests should not overlap; therefore, the Optical OFDM technique is used to provide orthogonality. 

\begin{figure}[H]
    \begin{center}
        \includegraphics[width=\linewidth]{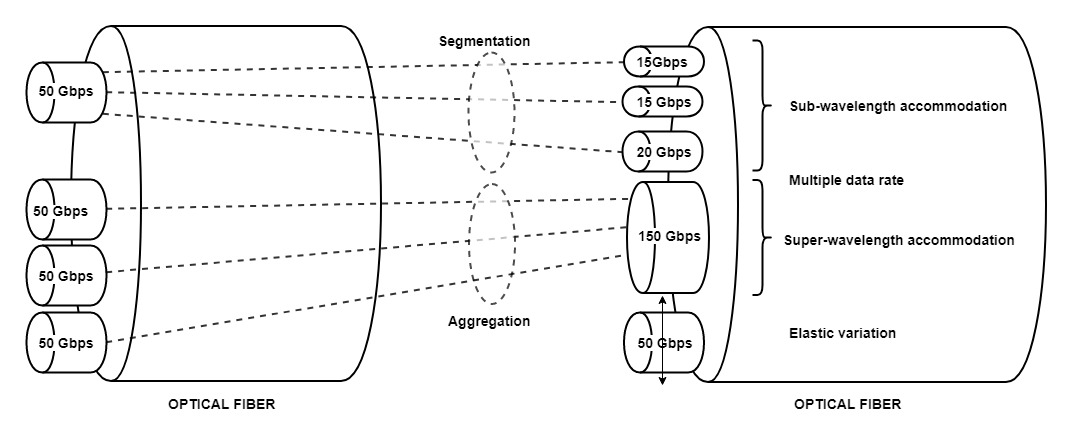}
        \caption{Characteristics of Elastic Optical Networks: Bandwidth Segmentation, Bandwidth Aggregation, Sub-wavelength and Super-wavelength Accommodation, and Energy saving \cite{EON1}.}
        \label{fig:EON2}
    \end{center}
\end{figure}

Compared to earlier optical networks, OFDM-based EON has various advantages such as more granular segmentation of bandwidth, the possibility of aggregation of bandwidth, for larger demands accommodation of multiple data rates as shown in Figure \ref{fig:EON2}, and Energy saving.

\begin{figure}[H]
    \begin{center}
        \includegraphics[width=\linewidth]{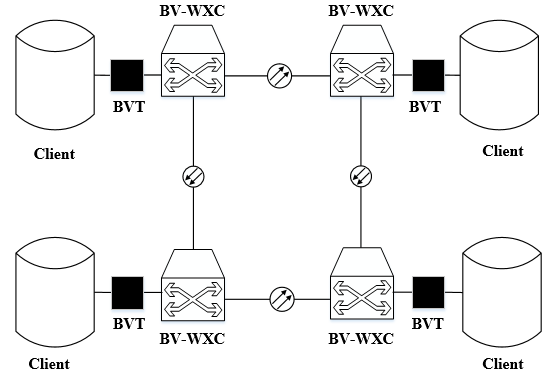}
        \caption{Architecture of EON with components such as Bandwidth Variable Transponder (BVT), and Bandwidth Variable-Wavelength Cross-connects (BV-WXCs).}
        \label{fig:EON3}
    \end{center}
\end{figure}

To support these characteristics of EON, three main components, used in the architecture of an elastic optical network, are Bandwidth Variable Transponder (BVT) or Sliceable Bandwidth Variable Transponder (SBVT), Reconfigurable Optical Add-Drop Multiplexer (ROADM), and Bandwidth-Variable Wavelength Cross-Connect (BV-WXC). BVTs allow the selection of slots by adjusting the modulation format. BVTs can trade spectral efficiency against transmission reach, e.g., for high-speed transmission, the Quadrature Amplitude Modulation (QAM) format can be used only for shorter distances. In comparison, for longer distances, BPSK or QPSK can be used. SBVTs are used for multi-variable optical flow (mixed line rates such as 200 Gbps, 300 Gbps and so on.). The ROADMs are used in conjunction with BVTs or SBVTs. The BV-WXC assigns a suitable size cross-connection with the appropriate spectral width supporting the flexible lightpath. It needs to flexibly customize its switching window according to the incoming optical signal spectral width. A typical arrangement of BVT and BV-WXC is shown in Figure \ref{fig:EON3}. 

Elastic Optical Networks function as a Circuit Switched Network with variable circuit capacity. For data transmission from Source Node to Destination Node, first, the connection is set up, and dedicated resources are allocated. The connection is released after the connection holding/service time is over. The resource allocation process has some constraints associated with it, due to functionality-related implications. Here, the resources are optical spectrum slots that must share boundaries with each other and should be same throughout the path, i.e., the Spectrum Contiguity and Spectrum Continuity Constraints have to be followed. The problem of finding the optimal path \footnote{The terms path and route are used interchangeably in this paper} and the optimal spectrum selection is called the Routing and Spectrum Assignment (RSA) problem.


\section{Routing and Resource Assignment}
Consider an \textit{N} node optical network. Suppose each node has \textit{N - 1} transceivers\footnote{transmitters (lasers) and receivers (photodetectors)}, and there are adequate resources on all the fiber links. In such a mesh network scenario, each node pair can be connected by a dedicated lightpath. Nevertheless, such a network design and resource assignment solution will result in overall high operational costs and may not be utilized to its fullest all the time.

To reduce the network cost, the traffic demand should be routed using minimum transceivers, wavelengths and switches with lesser number of ports\cite{RWA11}.

Routing and Resource Assignment problems need to be solved to satisfy the traffic bandwidth requirement while minimizing the overall resources to be consumed. It can be done using suitable routing algorithms while following the resource allocation constraints. Routing and Resource Assignment for a connection request is generally an NP-hard problem. It is called Routing and Wavelength Assignment (RWA) in Fixed-grid networks, and Routing and Spectrum Assignment (RSA) in Flexi-grid networks. Due to flexibility, the Flexi-grid adds an extra constraint to the RSA problem. Both fixed and flexible networks are circuit-switched optical networks, and hence, the resources are allotted at the connection setup and released only when the connection is dismantled.

    \begin{figure}[H]
    \centering
    \includegraphics[width=\linewidth]{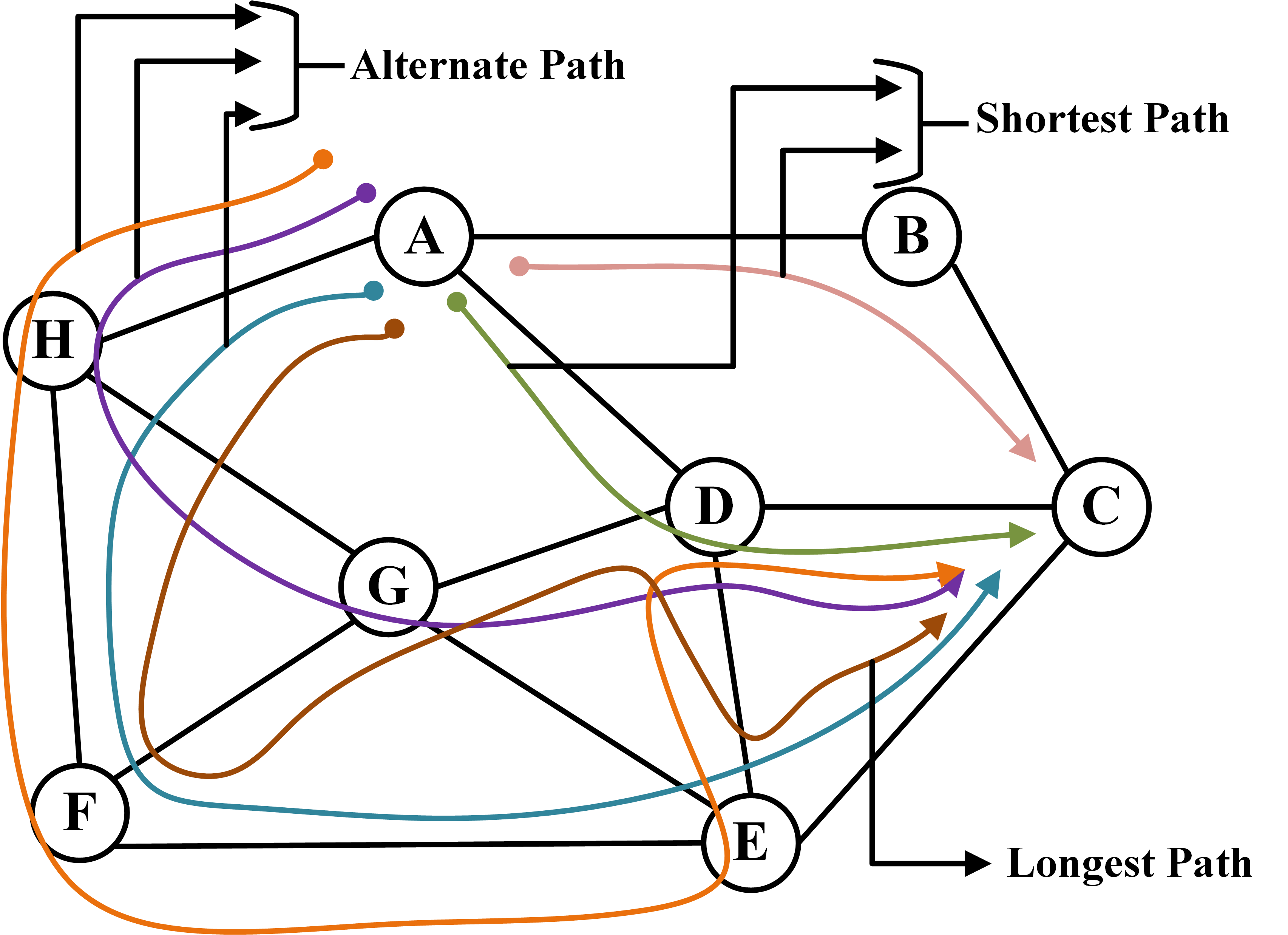}
    \caption{Different paths have been computed based on the number of hops from source node \textit{A} to destination node \textit{C}. Shortest Path: 2 hops route \textit{A-B-C} and \textit{A-D-C}, Alternate Path: Other than shortest path \textit{A-H-F-E-D-C} or \textit{A-H-G-D-C} or \textit{A-H-F-E-C}, and Longest Path \textit{A-H-F-G-D-E-C}. There are other paths also, but for simplicity, only few of them are computed.}
    \label{fig:path}
    \end{figure}

    \subsection{Routing}
    The goal of routing is to determine the optimum light paths between source and destination node pairs. Various routing methods have been studied in the literature. A few of them are explained below with the help of an example shown in Fig. \ref{fig:path}.

	\begin{enumerate}
		\item \textbf{Fixed Routing:} The fixed paths are pre-determined between pairs of nodes. A path can be determined using the Shortest Path Routing or Longest Path Routing Algorithms. The link parameters used to decide the route selection are usually static.
		\item \textbf{Fixed Alternate Routing:} \textit{k}-Shortest Path Routing or \textit{k}-Longest Path Routing can be used to find \textit{k} alternate paths between a pair of nodes. These paths may or may not be link-disjoint. 
		\item \textbf{Adaptive Routing:} In adaptive routing, the path between a pair of nodes can be selected based on the network's current state, as a result of some dynamic characteristics.
	\end{enumerate}

    \subsection{Resource Assignment}
    Depending upon the type of network, either Fixed-grid or Flexi-grid, the resource assignment problem is either a Wavelength Assignment or Spectrum Assignment problem, respectively.
    
    \begin{figure}
    \centering
    \includegraphics[width=0.8\linewidth]{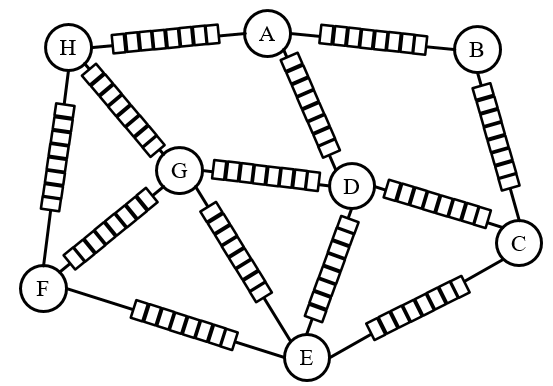}
    \caption{An optical network with \textit{8} nodes, \textit{13} bi-directional fiber edges and each edge consisting of \textit{8} slots.}
    \label{fig:wnet}
    \end{figure}
    	
    \begin{figure}
    \centering
    \includegraphics[width=0.8\linewidth]{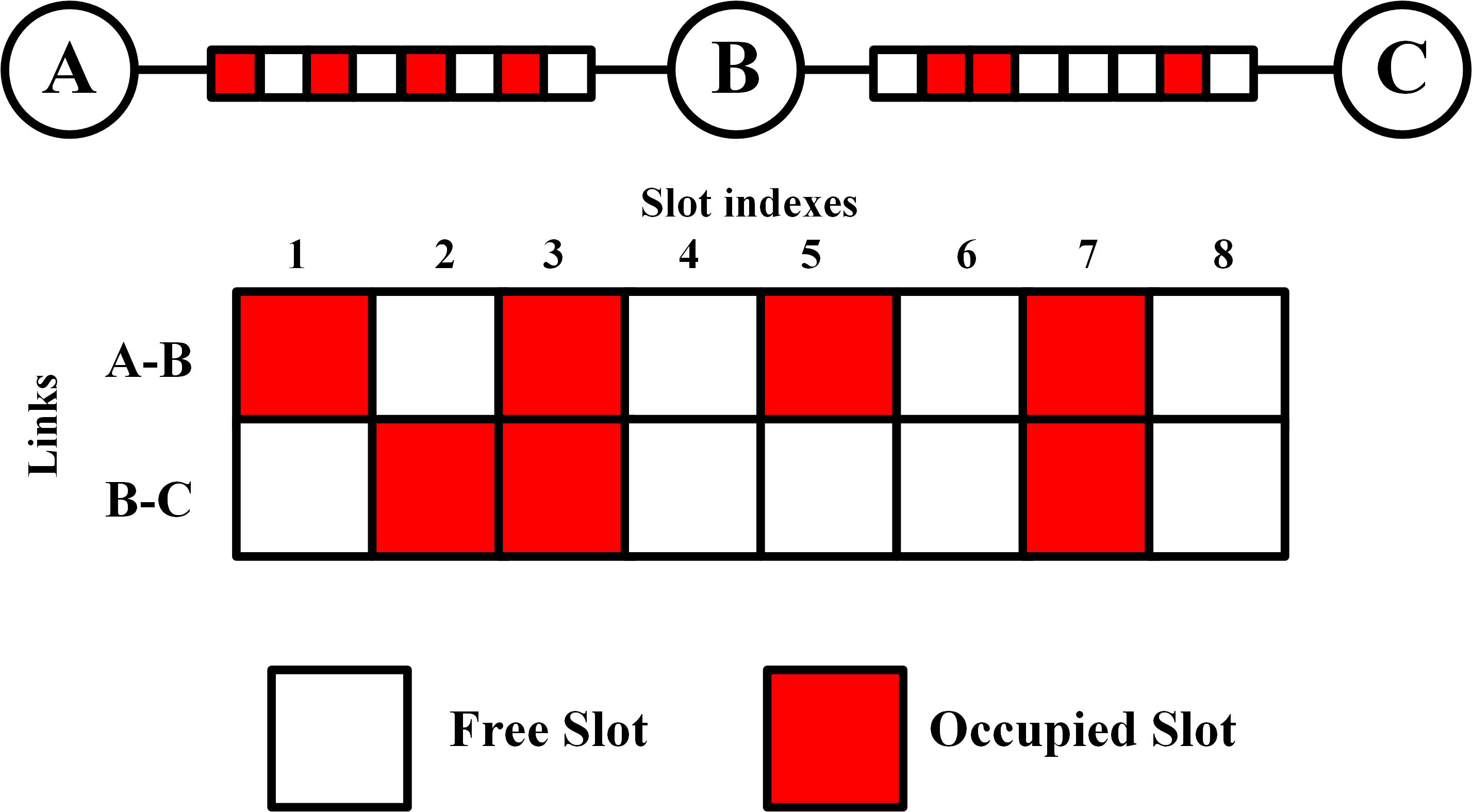}
    \caption{Current status of wavelength with free and occupied grids of a path in a network (fig. \ref{fig:wnet})}
    \label{fig:w1}
    \end{figure}

    \subsubsection{Wavelength Assignment}
\begin{figure}
    \centering
    \includegraphics[width=\linewidth]{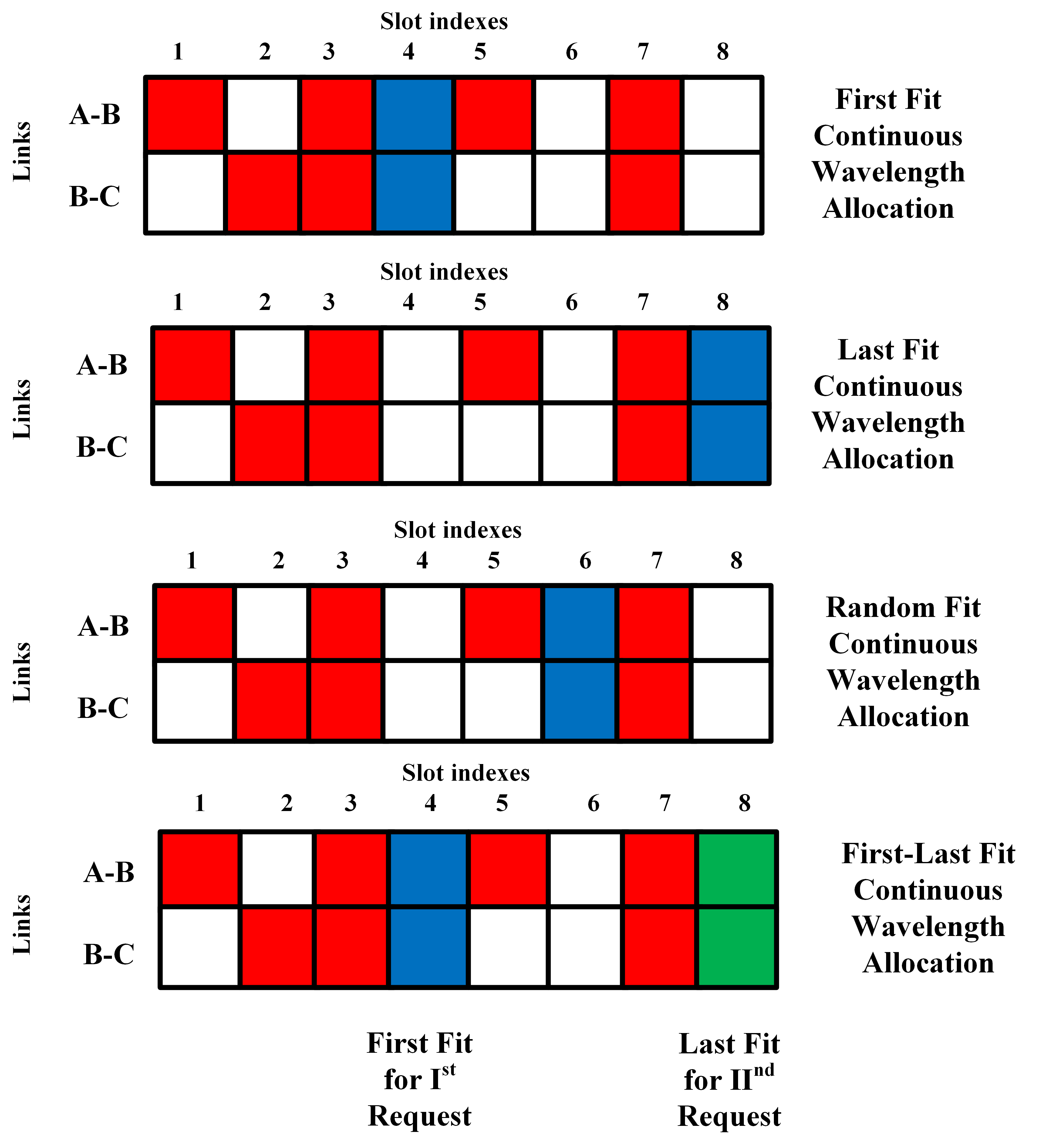}
    \caption{Continuous Wavelength Assignment Strategies: First Fit, Last Fit, Random Fit, and First-Last Fit. The wavelength allocated for the first and second lightpath requests are shown in this figure in blue and green, respectively.}
    \label{fig:wa}
\end{figure}

    The objective of the wavelength assignment is to allocate available wavelength slots\footnote{The wavelength allocated to a request is of a single unit, i.e., 50GHz\cite{RWAB}. Depending upon the type of modulation format, up to a certain amount of data can be accommodated.} to the incoming connection requests on the paths computed using the above routing algorithms while satisfying the following constraints. 
	\begin{itemize}
		\item \textbf{Wavelength Continuity Constraint:} The assigned wavelength index for a connection request must be same throughout the path, as shown in Fig. \ref{fig:wa}.
		\item \textbf{Wavelength Non-Overlapping Constraint:} The assigned wavelength indexes cannot be same for any two different connection requests passing through same link. 
	\end{itemize}

	The wavelength grids are indexed and a catalogue of used and available indexes are maintained.
	 
	\begin{enumerate}
		\item \textbf{First Fit:} This strategy often selects the lowest available wavelength index and assigns them to the lightpath request, as shown in fig. \ref{fig:wa}.
		\item \textbf{Last Fit:} This strategy often selects the highest available wavelength index and assigns them to the lightpath request, as shown in fig. \ref{fig:wa}.
		\item \textbf{Random Fit:} This strategy often selects the available wavelength index randomly and assigns them to the lightpath request, as shown in fig. \ref{fig:wa}. 
		\item \textbf{First-Last Fit:} This strategy selects the lowest available wavelength index for odd-numbered connection requests and the highest available wavelength index for even-numbered connection requests and assigns them to the lightpath request as shown in Fig. \ref{fig:wa}. It alternates between the first fit and the last fit for the arriving connection requests to maximize the chances of finding unused wavelength indices.
	\end{enumerate}	
	
    \subsubsection{Spectrum Assignment}
    The objective of spectrum assignment is to allocate available spectrum slots to the path computed using the routing algorithms while satisfying the following set of constraints\cite{ch21}:
	
	\begin{itemize}
		\item \textbf{Spectrum Contiguity Constraint:} The assigned spectrum slots indexes for a connection request must share boundaries with each other as shown in Fig. \ref{fig:cc}.
		\item \textbf{Spectrum Continuity Constraint:} The assigned contiguous spectrum slots indexes for a connection request must be same throughout the path as shown in Fig. \ref{fig:cc}.
		\item \textbf{Spectrum Non-Overlapping Constraint:} The assigned slot indexes cannot overlap with one other; that is, in any link, no two connection requests will be assigned same spectrum slots. 
	\end{itemize}
	
	\begin{figure}
    \centering
    \includegraphics[width=0.8\linewidth]{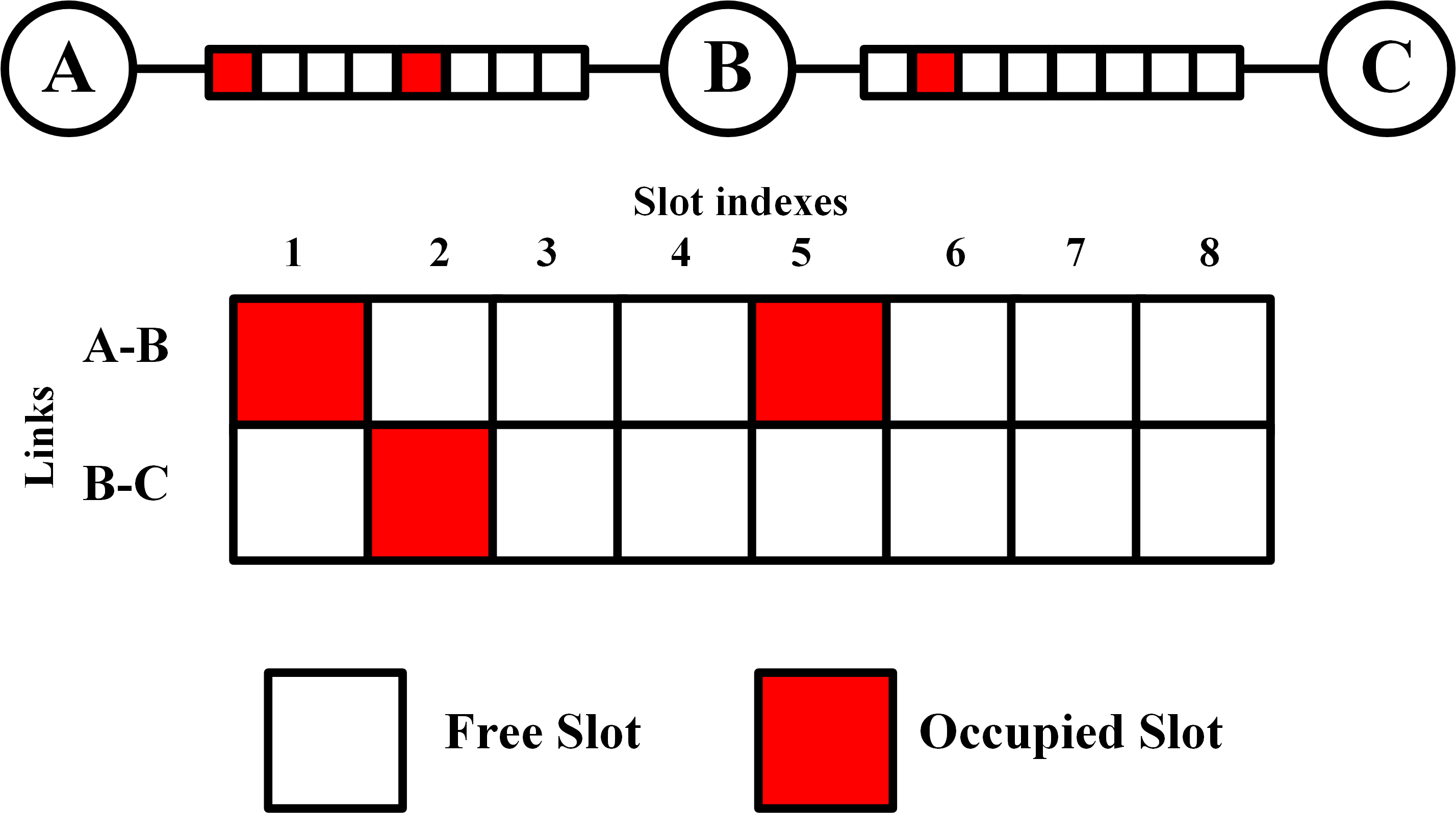}
    \caption{Current status of 12.5GHz spectrum slots with free and occupied slots on a path for a network (fig. \ref{fig:wnet})}
    \label{fig:w1}
\end{figure}

\begin{figure}
    \centering
    \includegraphics[width=0.8\linewidth]{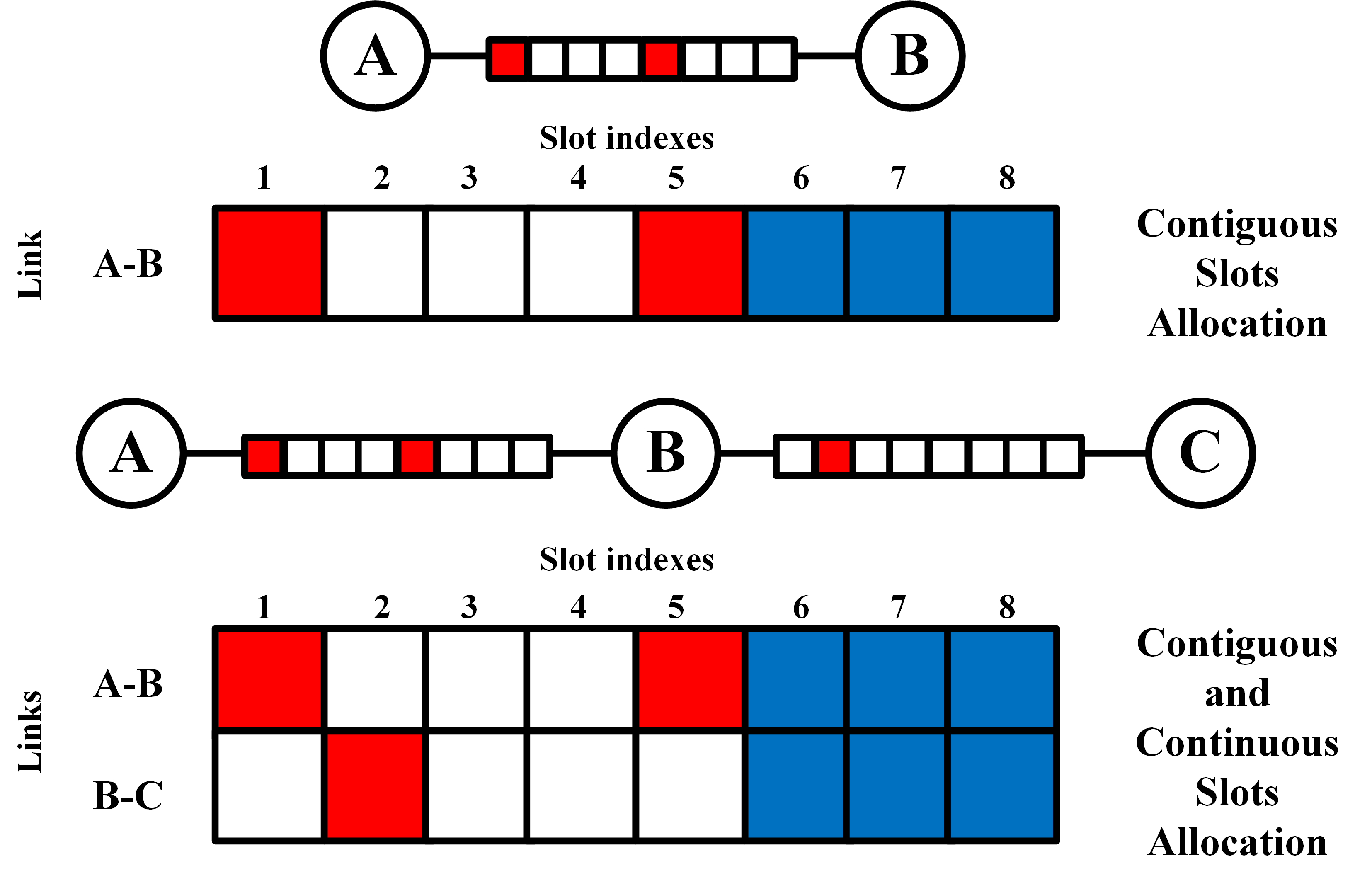}
    \caption{Spectrum Assignment Constraints: Contiguous and Continuous Spectrum Slots Assignment. Three slots connection assignment.}
    \label{fig:cc}
\end{figure}

\begin{figure}
    \centering
    \includegraphics[width=\linewidth]{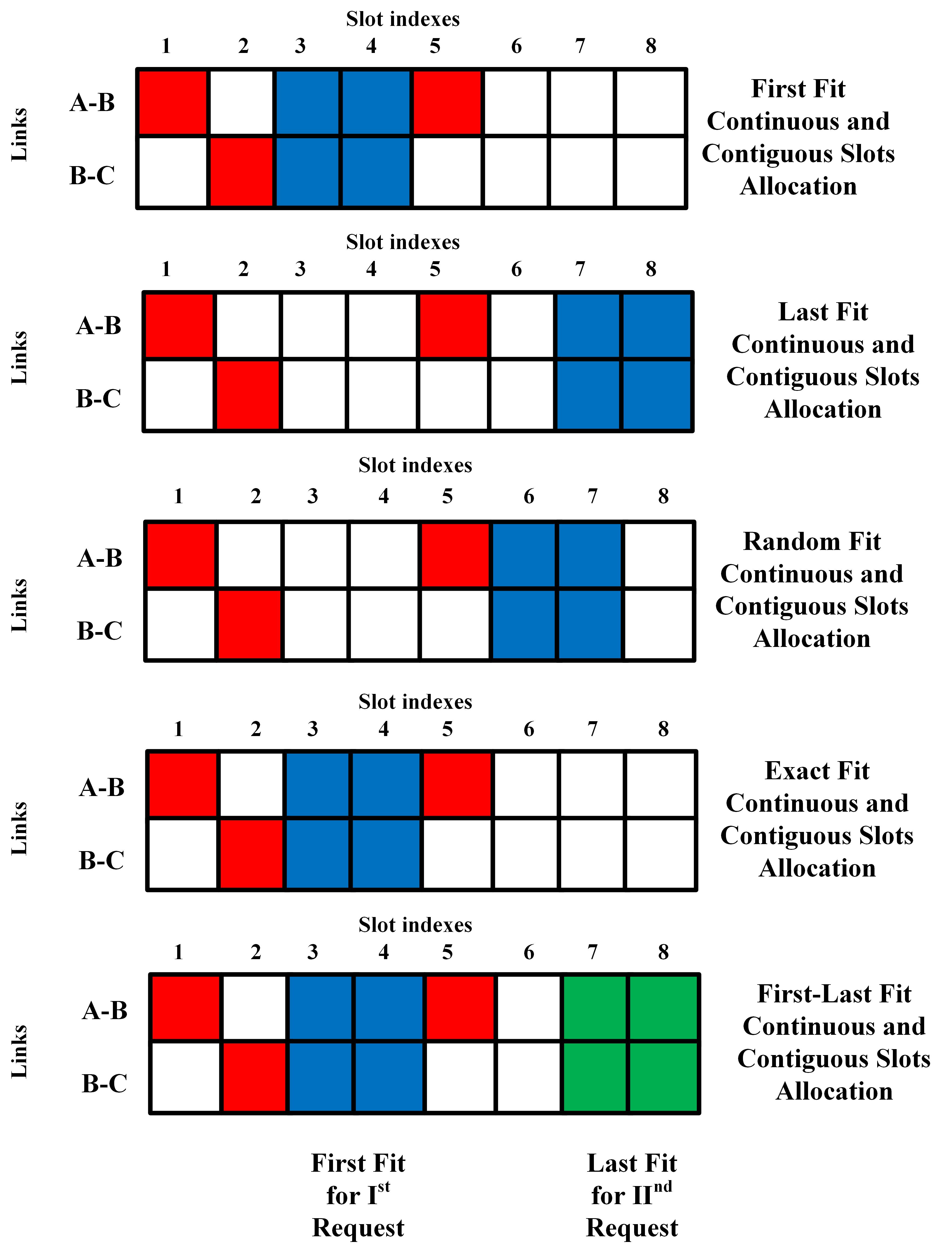}
    \caption{Contiguous and Continuous Spectrum Slots Assignment Strategies: First Fit, Last Fit, Random Fit, Exact Fit, and First-Last Fit.}
    \label{fig:s2}
\end{figure}

The spectrum grids are also indexed, and a catalogue of used and available indexes is maintained just like for wavelength grids. The purpose of these spectrum grids is to help in visualizing the network spectrum status.
	\begin{enumerate}
		\item \textbf{First Fit:} This strategy selects the lowest available indexed spectrum slots and assigns them to the lightpath request.
		\item \textbf{Last Fit:} This strategy selects the highest available indexed spectrum slots and assigns them to the lightpath request.
		\item \textbf{Random Fit:} This strategy randomly selects from all the available indexed spectrum slots and assigns them to the lightpath request.
		\item \textbf{Exact Fit:} This strategy finds all available indexed spectrum slots such that adjacent slots on both sides are occupied. If more than one option is there, then either the first fit or the last fit is used among them to choose the group of slots. If no exact fit is found, the first or last fit, as the case may be, is used.
  
		\item \textbf{First-Last Fit:} This strategy selects the lowest available indexed slots for odd-numbered connections and the highest available indexed slots for even-numbered connections and assigns them to the lightpath requests. Alternatively, the lowest available indexed slots for connections with an odd number of required slots and the highest available indexed slots for connections with an even number of required slots are found and assigned to the lightpath request.
	\end{enumerate}
	

The problem of Routing and Wavelength Assignment (RWA) in fixed grid optical networks was first introduced by Chlamtac \textit{et al.}\cite{RWA1} in 1992 where the heuristics for centralized and distributed dynamic lightpath establishment was given. Kelly {\em et.al.} \cite{RWA2} proposed a reduced load approximation model for a circuit-switched scheme. The extended version was later proposed by Chung {\em et.al.} \cite{RWA3} where an analytical model of wavelength-routed optical network for the single-hop route was discussed. For at most three hops route, the reduced load approximation was discussed in \cite{RWA4}. \cite{RWA14} uses the reduced load approximation scheme and proposed Independence Model and Correlation Model for arbitrary traffic and topologies. Various heuristics have been proposed for solving the RWA problem, and simulations have quantified their performance for different network topologies in \cite{RWA5} \cite{RWA6}, \cite{RWA7}, \cite{RWA11}. However, to reduce the blocking of lightpath requests and to increase the utilization of optical fiber bandwidth, the use of wavelength converters was suggested. The analytical modelling of RWA with wavelength converter placement problem has been studied in \cite{RWA8}, \cite{RWA9}, \cite{RWA10}, \cite{RWA12}, \cite{RWA13}, \cite{RWA15}.

In fixed bandwidth channels (i.e., fixed spectrum grid network), when a lightpath of a smaller bandwidth is required, the bandwidth is wasted. Further, the lightpaths requiring larger bandwidth than the single channel capacity, cannot be set up. Flexible grid optical networks were proposed to allow for smaller or larger bandwidth without wasting bandwidth thereby achieving higher spectrum utilization. With a flexible grid network, the RWA problem becomes RSA (routing and spectrum assignment) problem. The research community has addressed RSA using a number of methods and employing different routing schemes, spectrum assignment algorithms based on the cost of the path, by using distance-adaptive modulation level manipulation. The concept of Elastic Optical Network was introduced by Jinno \textit{et al.} \cite{EON1}. They proposed that according to the required bit rate of a connection request, a group of frequency slots (FSs) can be allocated based on fixed-alternate routing and first-fit frequency assignment. A heuristic-based approach was adopted to show that low spectrum usage is possible by using the distance adaptive schemes while remaining within the target BER bounds. Gerstel {\em et.al.} \cite{RSA4} and Lohani {\em et. al}\cite{RSA9} gave overview of such Elastic Optical Networks. In \cite{RSA2}, the concept of modulation format was introduced in Elastic Optical Network. In \cite{RSA3}, \cite{RSA5}, \cite{RSA6}, \cite{RSA7}, \cite{RSA8}, \cite{RSA10}, the problem of Routing and Spectrum Allocation in Flexi-grid networks was investigated. In these, various heuristics for RSA were proposed for tackling the problem of fragmentation in the optical spectrum. 

Dallaglio \textit{et al.}, \cite{RSA6},  proposed a dynamic routing, spectrum and transponder assignment (RSTA) scheme that unified transponder selection with RSA. The constraints for transponder selection were also integrated with other constraints. The scenarios of using a Multi-Wavelength Sliceable Bandwidth Variable Transponder (MW-SBVT) in place of a Multi-Laser SBVT (ML-SBVT) was also considered to get maximum benefit from both the variable transponder technologies. It is also suggested that MW-SBVT cab be more suitable for high-bitrates super-channels. There are a few works, which consider multipath routing for RSA in EON to accommodate more number of connection requests, or to lower the blocking probability. 

In \cite{c6}, Zhou \textit{et al.} proposed an efficient genetic algorithm to solve dynamic routing, modulation and spectrum assignment (RMSA) for EONs. The algorithm is designed for multi-objective optimization. However, genetic algorithm-based methods cannot guarantee optimality and solution quality deteriorates with an increase in problem size.

Klinkowski {\em et.al.} \cite{c7} formulated RSA as an ILP, having a predefined set of paths, with the objective to minimize the number of frequency slots allotted to at least one traffic demand in the spectrum. They faced the main difficulty due to a large number of frequency slots, resulting in a huge number of constraints. A heuristic-based Adaptive Frequency Assignment- Collision Avoidance (AFA-CA) was further proposed where demands are processed adaptively, i.e., the next demand is selected according to previously allocated demands as well as the current usage of frequency slots.

Castro {\em et.al.} \cite{c8} proposed heuristics for Dynamic routing and spectrum (re)allocation, where the allocated lightpath requests are reallocated optical spectrum such that there will be room for new connection requests.

In \cite{c9}, Chen {\em et.al.} proposed a distance-adaptive modulation format-based multipath routing scheme, and applied it to dynamic traffic scenarios. The blocking probability is significantly reduced using this method while using comparable spectrum resources as in single path routing. However, guard band size also played a crucial role here. It is highlighted in this work that the multipath routing seems to be more suitable when the guard band size is small, and the traffic load is high.

\subsubsection*{Problems due to Constraints}
  
\begin{figure}
    \centering
\includegraphics[width=0.8\linewidth]{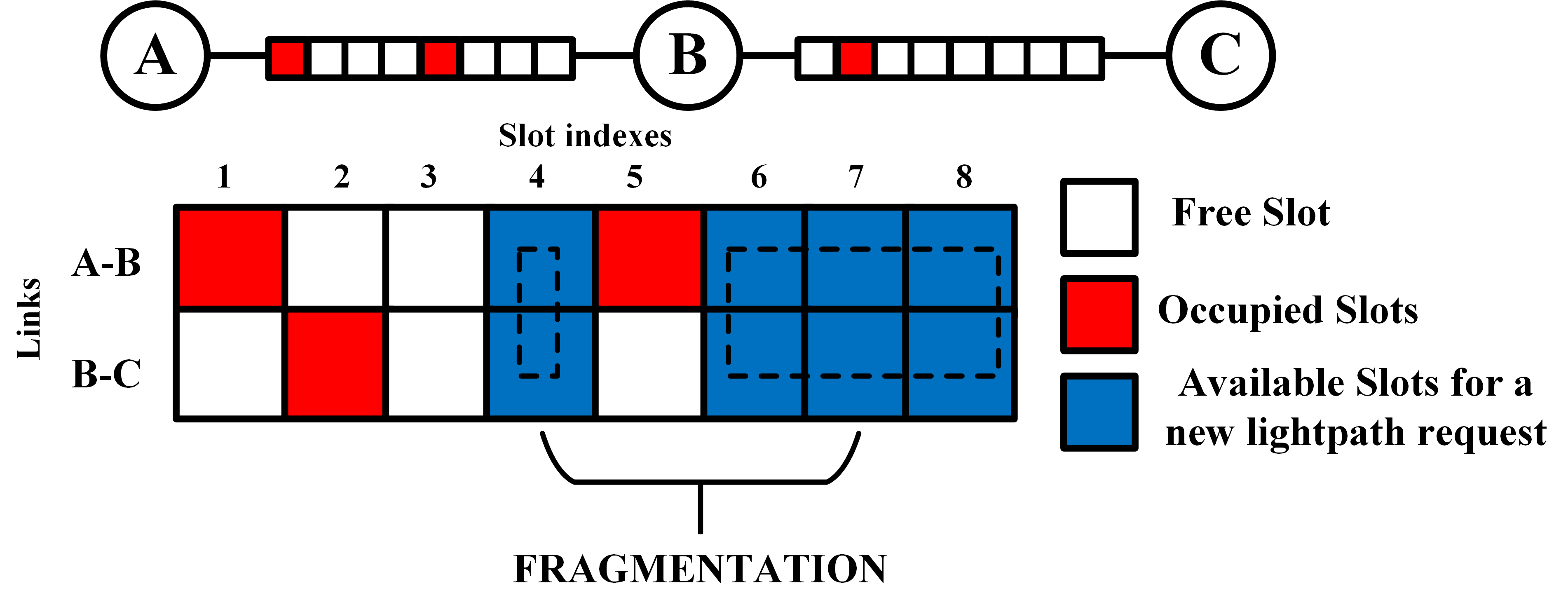}
    \caption{Fragmentation problem due spectrum assignment constraints.}
    \label{fig:frag}
\end{figure}

		As the networks operate with dynamic and heterogeneous traffic scenarios, continuous setting up and dismantling of the lightpath requests with strict adherence to the spectral assignment constraints lead to the problem of \textit{``Fragmentation"} in the optical spectrum. Fragmentation is the scenario when a few spectral resources are available for assignment yet cannot be used as per the constraints. Fragmentation is illustrated with the help of an example shown in Fig. \ref{fig:frag}. A lightpath request arrives from source node \textit{A} destined to node \textit{C} with the requirement of four slots. Although the four slots are available, they cannot be allocated due to their availability at different spectrum indices in the links of the path, thus their non-compliance with RSA constraints. As we have to follow spectrum assignment constraints, the connection request is blocked. Fragmentation management becomes crucial in improving the overall lightpath request admission in the network by minimizing or mitigating fragmentation in the network spectrum. \\
     
There are many works in the literature related to the measurement of fragmentation level and then using the defined metrics in making routing decisions. The contiguity of the spectrum slices is one of the most prominent indicators of fragmentation level. In \cite{RSA11}, Chatterjee {\em et.al.} have extensively studied the various forms of fragmentation. This work presented the types of fragmentation metrics and defragmentation strategies, and non-defragmentation strategies. In \cite{RSA12}, Pederzolli {\em et.al.} have presented one of the many methods to calculate fragmentation level and then employing it in RSA decision-making. In \cite{VFM}, authors presented a Vectored fragmentation metric consolidating the fragmentation due to continuity and contiguity constraints, in a single indicator. Yin {\em et.al.} \cite{RSA13} and Zhu {\em et.al.} \cite{RSA14} have explored fragmentation-aware routing and spectrum allocation in their works. Various parameters of the spectrum allocation in the links of the network are observed, and based on them, routing decisions are made. These parameters may or may not directly affect the fragmentation level. Kim {\em et.al.} \cite{RSA15} gave an analytical model of fragmentation in the optical spectrum for two-services \footnote{Two-services means two sets of slots requirement.}.
   
        Intuitively, if the spectrum assignment constraints are relaxed, then the problem of fragmentation could be solved. If a spectrum band converter is present at each node, the fragmentation can be avoided by relaxing the continuity constraint, but such converters are expensive to deploy. If no such measures are there and fragmentation has also occurred in the spectrum, then retorting by using various fragmentation management protocols could help in mitigating the spectrum's fragmentation. These are broadly classified into non-de-fragmentation and non-defragmentation types. De-fragmentation technique focuses on re-configuring the resource allocation of the active connection requests, i.e., a new set of spectrum slots or a new set of path-spectrum slots are assigned. De-fragmentation protocols are further classified into hitless and non-hitless, depending on the disruption of the active connections in the requests, for some time, while performing rearrangements. Due to added rearrangement step, de-fragmentation approaches are complex and costly. On the other hand, non-defragmentation protocols do not affect the existing connections, and preparatory measures are taken to avoid fragmentation. 
        
        To reduce the calculation overheads, defragmentation could be performed periodically. The trigger point could be the number of arrived connection requests, or threshold-based fragmentation indicators. Using fragmentation level indicators could be advantageous, as the defragmentation is triggered only when necessary. Defragmentation can be Proactive or Reactive depending on periodicity, \cite{DF6}, or triggering due to blocking of certain connection requests in recent past, respectively. Authors in \cite{DF1} have studied the conditions in which the reactive and proactive approaches can outperform each other. A number of notable works explore the de-fragmentation techniques considering the appropriate initiating mechanism while maintaining the resource utilization efficiency at an acceptable level \cite{DF2}-\cite{DF8}.

%

\section{Survivability against failure}
Other than Routing and Resource Assignment, Resilience or Survivability is a very widely researched area in optical networks. Optical networks need to carry an enormous amount of traffic while maintaining service continuity even in the presence of faults. Failure of even a single link\footnote{span, link and edges are used interchangeably in this paper} will result in loss of a substantial amount of data if not protected automatically. Therefore, survivability against link or path failures is an essential design requirement for high-speed optical networks. The goal of a survivability scheme is to offer reliable services for large volumes of traffic even in the presence of failures\footnote{Fiber cut, human-made errors or natural disasters such as earthquakes, hurricanes, etc.} as well as other abnormal operating conditions \cite{PR}.

\begin{figure}
    \centering
    \includegraphics[width=\linewidth]{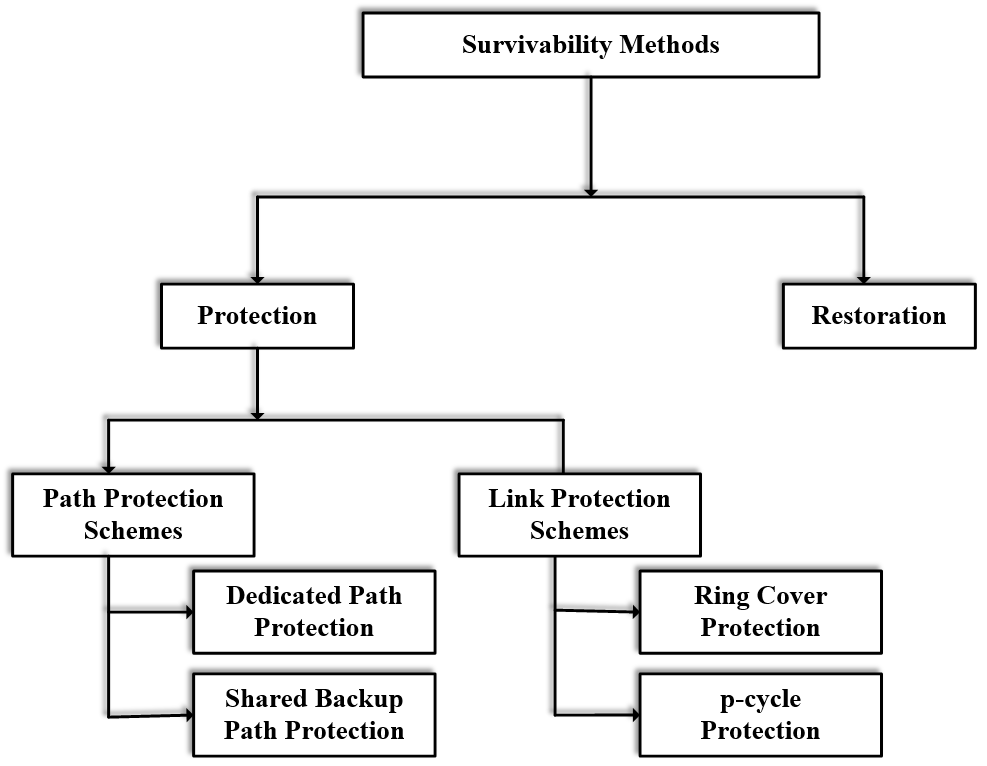}
    \caption{Survivability Mechanisms}
    \label{fig:hierarchy}
\end{figure}

Network survivability implies continuous provisioning of services even in the presence of failures. The two mechanisms needed in any survivability implementation are protection and restoration. The hierarchy of survivability methods is shown in fig.\ref{fig:hierarchy}.

\subsection{Protection}
Protection is a technique in which spare capacity (if it exists) is allocated a priori to restore the specific links or paths for working connection whenever a failure occurs. In an optical network, usually, a pair of fibers are allocated for each link in each direction. One of the fiber is called working fiber and the other one is protection fiber. Alternatively, the wavelengths can be working or protection wavelengths. Under normal conditions, data is transmitted through working fiber/ wavelengths. If the working fiber fails, the transmitting node switches its data from the working fiber/wavelengths to the protection fiber/wavelengths in other links which form the protection. Protection paths are generally pre-computed for faster restoration and higher reliability. Optical layer protection is classified as Path Protection and Link Protection.
    
\subsubsection{Path Protection Schemes}

In path protection, the entire lightpath from a source node to a destination node is protected by an alternate path between same pair of nodes. The alternate path may not share any link (for link failure protection) or node (for node failure protection) with the working path. Usually, the main working path is called the primary path, and the alternate path is called secondary or backup path. Under normal conditions, high-priority traffic flows through main working fiber/wavelengths and low-priority traffic flows through the secondary (protection) path. The low-priority traffic is dropped whenever a failure occurs, and the high-priority traffic is switched over to the link (and node) disjoint protection path. Fault localization is not required because protection path is link-disjoint. This protection is further classified as Dedicated Path Protection and Shared Backup Path Protection.

\subsubsection*{Dedicated Path Protection (DPP)}
In dedicated path protection, the working path has a dedicated\footnote{The backup path is not shared with any other working path.} link and node disjoint backup path for protection against link failures or node failures. When there is no failure, the backup path carries the low-priority traffic for efficient capacity utilization. Whenever there are failures in the working path, the low-priority traffic in the backup is dropped, and the traffic in the working path is switched over to the backup path. Due to dedicated protection for each working path, more resources are required to protect the entire traffic. To reduce the spare capacity requirement, a protection path can be shared with multiple working paths. It is called shared backup path protection.

\begin{figure}
    \centering
        \includegraphics[width=0.8\linewidth]{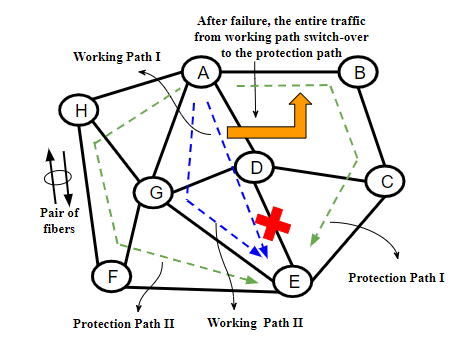}
        \caption{Dedicated Path Protection: The protection path from \textit{A-B-C-E} protects only one working path i.e., \textit{A-D-E} not \textit{A-G-E}. For \textit{A-G-E}, there is another protection path i.e., \textit{A-H-F-E}}
        \label{fig:dpp}  
\end{figure}

\subsubsection*{Shared Backup Path Protection (SBPP)}
In SBPP, a single backup path protects the number of primary working paths. When a failure occurs in any one of the primary paths, then the failed path traffic is switched to the pre-configured backup path. The primary paths must be link and node disjoint with the backup path as well as among themselves. The main advantage of SBPP scheme is its resource efficiency. However, its switching speed is not as good as that of the p-cycles\footnote{one of the link protection technique}.

\begin{figure}
    \centering
       \includegraphics[width=0.8\linewidth]{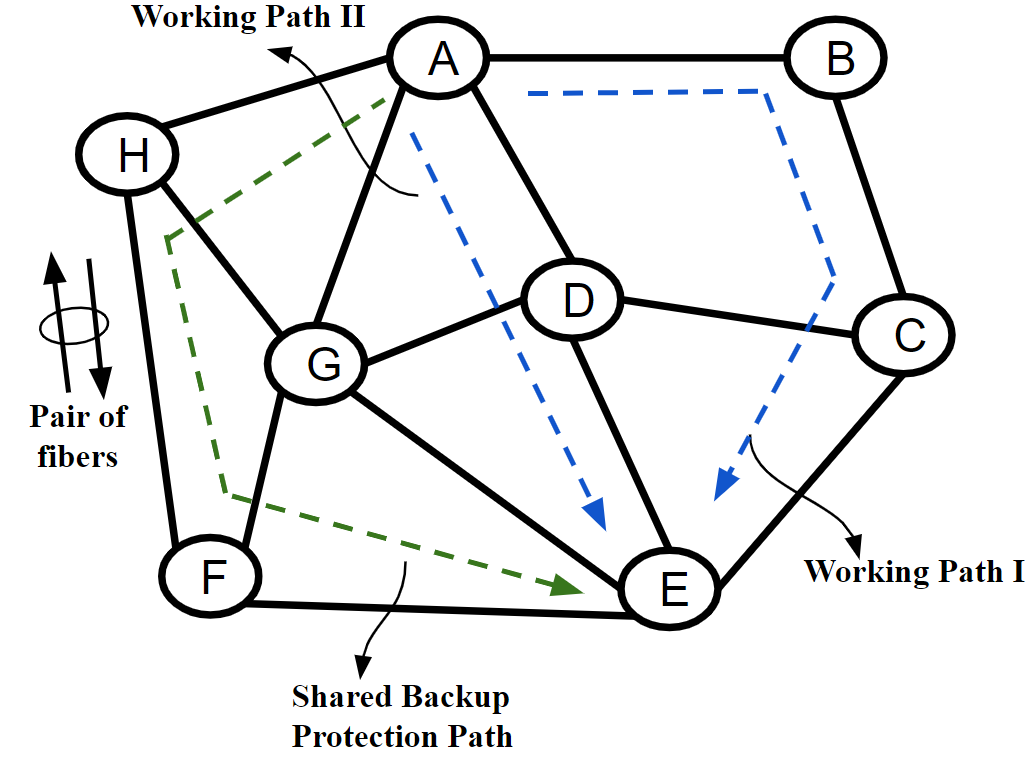}
       \caption{Shared Backup Path Protection: A backup path \textit{A-H-F-E} is shared among two working paths i.e., \textit{A-D-E} and \textit{A-B-C-E}}
       \label{fig:sbpp}  
\end{figure} 

\subsubsection{Link Protection Schemes}
\begin{figure}
    \centering
        \includegraphics[width=\linewidth]{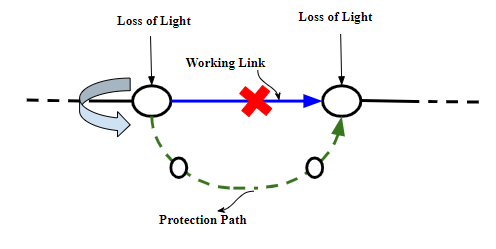}
        \caption{Fault localization in Link Protection.}
        \label{fig:fl} 
\end{figure}

In a link protection scheme, instead of switching over the entire traffic from primary path to another backup path, the traffic through faulty link is restored via another path joining the two endpoints of the link \cite{link}. The other links on that primary path remain as it is. In this strategy, every link is provisioned with a backup path. The link protection is shown in fig.\ref{fig:fl}. 

The switching happens whenever there is a loss of light at the end nodes of the link. The link protection schemes are fast because of faster fault localization. Two of the link protection schemes are Ring cover and p-cycles.

\subsubsection*{Ring Cover}
\begin{figure}
    \centering
        \includegraphics[width=0.8\linewidth]{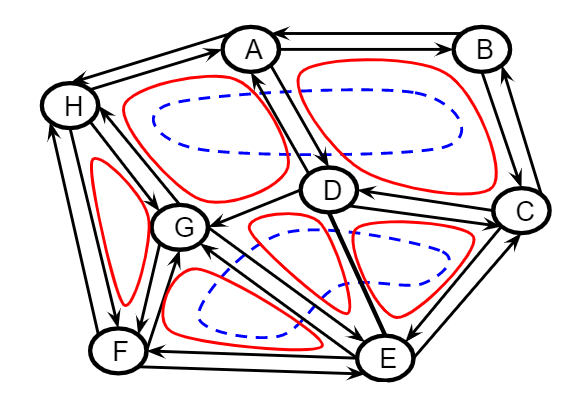}
        \caption{Ring Cover Protection}
        \label{fig:rc}   
\end{figure}

In this scheme, cyclic paths are identified in the network such that each edge is traversed by at least one cycle. These cyclic paths are called ring covers \cite{RC1}. In this scheme, some of the links may be covered with more than one cycle, which results in additional redundancy. In fig.\ref{fig:rc}, few of the cycles are shown. Fiber links are used to form these ring covers (cycles). 

Every cycle requires four fibers (two fibers for the bi-directional working path and the remaining two for bi-directional backup path). If $N$ rings cover an edge, the edge will need $4N$ fibers. A working path is setup by traversing the link through working fibers in the cycle. When a link fails, the traffic in the working fiber on the link is switched over to protection fiber of the cycle. It is desirable to reduce the amount of fiber used for forming the cycles while maintaining the desired resilience.

\begin{figure}
    \centering
        \includegraphics[width=0.8\linewidth]{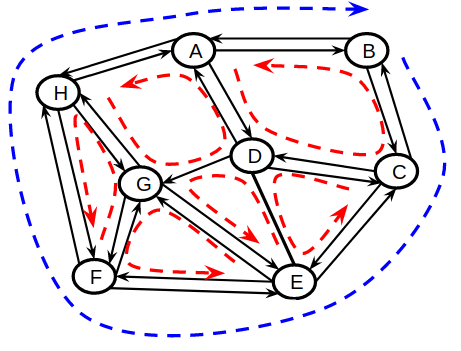}
        \caption{Double Cycle Ring Cover Protection}
        \label{fig:dcrc}   
\end{figure}

To minimize spare capacity, the double-cycle ring cover was proposed. In the double-cycle ring cover, the graph with bi-directional working links and protection links is considered. Each edge is provided with protection by two unidirectional cycles each one in opposite direction on the common link \cite{RC2}. Fig.\ref{fig:dcrc} show such a configuration. The directed cycles are chosen such that all of them are anticlockwise inside the graph. For the areas which is outside the graph, a clockwise cycle is chosen. Each link is now protected by two different cycles in each direction. The spare capacity reduces to $2N$ in this configuration. There are no cycles formed with working capacity as in Ring Cover protection. One of the drawbacks of double cycle ring covers is different signal delays in forward and backward directions after the restoration of traffic passing through the failed link. Further, the restoration times are also different in the two directions.

\subsubsection*{{p}-cycle (pre-configured cycle)}
In ring cover and double cycle ring cover, only on-cycle failures are restored. If the straddling link\footnote{only end nodes are the part of the cycle, and not the link} failures can also be restored by the cycle, then we can further reduce the spare capacity requirement. This kind of configuration is called p-Cycle based protection mechanism. The concept of p-cycles was first introduced by Grover {\em et. al.} \cite{p1}. p-Cycles in an optical mesh network provide same switching speed as the ring-based protection $(< 50ms)$\footnote{Bi-directional Line Switched Ring restoration speed} and capacity efficiency as in mesh networks \cite{p1}, \cite{p2}.

\begin{figure}
    \centering
        \includegraphics[width=0.8\linewidth]{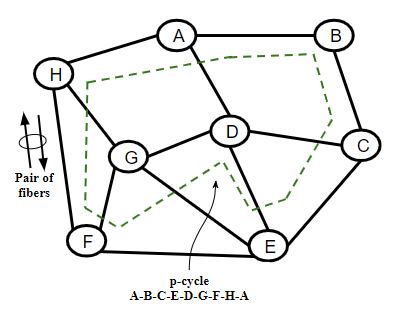}
        \caption{A p-cycle, \textit{j}}
        \label{fig:p1}
\end{figure}
        
\begin{figure}
    \centering
        \includegraphics[width=0.8\linewidth]{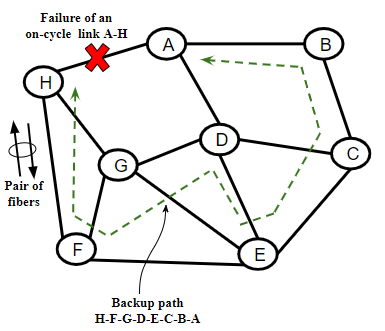}
        \caption{A failed on-cycle link \textit{i}, p-cycle \textit{j} provides single unit of protection path}
        \label{fig:p2}
\end{figure}

\begin{figure}
    \centering
        \includegraphics[width=0.8\linewidth]{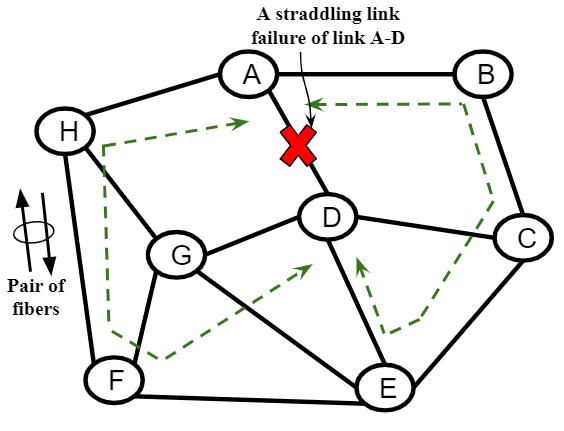}
        \caption{A failed straddling link \textit{i}, p-cycle \textit{j} provides two units of protection path}
        \label{fig:p3}
\end{figure}

Whenever there is a link failure, each p-cycles based on whether the failed link is on-cycle or straddling to the p-cycle, will provide a single unit (on-cycle) or two units (straddling\footnote{A straddling link must have its edge nodes on the p-cycle, but it's not part of the cycle.}) of protection to working capacity as shown in fig.\ref{fig:p1}, \ref{fig:p2}, and \ref{fig:p3}.

\subsection{Restoration}
Restoration is the almost real-time switch-over to the protection paths/links on the occurrence of a failure. Restoration paths are calculated after the failure if the protection is not provisioned. Therefore, the switching speed of the restoration mechanism without pre-provisioned protection is slow. Also, $100\%$ survivability is not guaranteed.

\subsection{Multiple Failure Survivability}
Survivability schemes can also be classified based on the number of concurrent faults they are designed to protect. Schemes encompassing multiple simultaneous protection provide networks greater resilience but at the expense of more network resources. In order to design optical networks capable of handling higher number of simultaneous faults, not only the required resources but also the complexity of the optimization problem rises. Thus, optical networks are designed to endure failures that are most likely to occur. Even though a great deal of research is focused on single-link failure solutions, simultaneous two-link protection strategies have also been investigated.

Protection schemes designed to address single-link failures can extend their protection against certain double-link failures as well, providing resilience to a higher number of faults besides consuming fewer resources. Availability metrics, that evaluate the duration during which demand or service remains operational in an optical network have been studied in the context of multi-link protection \cite{avl1, avl2, avl3, avl4}. The literature explores various availability metrics including Non-restored Working Capacity (NWC), Restorability (R), and Service Path Unavailability (SPU). In a p-cycle based survivability scheme \cite{Schupke}, Schupke et al observed that maximizing p-cycle increased restorability while increasing the capacity requirement. On the other hand, minimizing the maximum working capacity coverage of p-cycles is capacity efficient and gives better restorability. In \cite{athe3}, impact zone has been utilized to assess the reduction in protection capacity resulting from a link failure when employing a shared protection. These studies can be used to improve the protection capability of optical networks by identifying and optimizing the factors that affect survivability.

A few novel approaches have also been explored by the researchers for two-link failure protection. The pre-Configure Polyhedron(PCP) \cite{PCP} aims to protect networks from multi-link failures but requires searching for backup paths within the PCP before switching. Another approach pre-Configured Ball(p-Ball) \cite{PBall}, minimizes backup links for dual failure restoration using a ring set but increases restoration time due to a routing algorithm. In \cite{tset}, the Tie-set Based Fault Tolerance(TBFT) method autonomously finds tie-sets covering all network edges to recover double-link failures. Zhang at al. \cite{Pkecks} proposed pre-configured k-edge-connected structures(p-Kecks) for protection from multi-link failures in optical networks, forming subgraphs with k-link-disjoint paths between any two nodes. These schemes not only demand higher capacity but also present increased complexity in capacity optimization. In \cite{athe2}, Athe et al introduce p-cycle-based schemes for dual link protection with a focus on enhancing capacity efficiency and reducing the complexity of the optimization problem.

Natural disasters can significantly impact telecommunication systems, making the study of disaster survivability a critical area of focus \cite{natdisas}. Disaster survivability models can be categorized into deterministic and probabilistic approaches. The deterministic approach \cite{determ} assumes that failure occurs with equal probability within a defined disaster zone, whereas the probabilistic approach \cite{prob1, prob2} takes into account the probability of failure based on intensity and proximity.

Numerous strategies have been proposed to mitigate the multiple failures resulting from disasters. One approach involves representing disaster regions as Shared Risk Link Groups (SRLG) and subsequently assigning separate working and backup paths to the links in the network \cite{srlg}. Also, multipath routing has been explored, where data is divided and sent over multiple disjoint paths to speed up the transmission \cite{multp}. The allocation of protection capacity to mitigate the impact of multiple failures stemming from natural disasters may entail substantial costs. Consequently, determining the degree and scope of protection for disaster survivability are critical aspects of consideration.
%
\subsection{Related Work}
Survivability can be incorporated into the networks either by allocating redundant capacity before the occurrence of failure or by allocating available capacity after the occurrence of failure. Protection is a technique in which spare capacity (if it exists) is allocated a priori to restore the working connection whenever a failure occurs. For protection in an optical network, usually a pair of fibers are allocated for each link in each direction. One of the fiber is called working fiber and the other one is protection fiber. Protection paths are generally pre-computed for faster restoration while providing higher reliability. The protection provided to paths is further classified as Dedicated Path Protection \cite{DPP} and Shared Backup Path Protection \cite{SBPP}.

Other than path protection, there is a link protection scheme, where instead of switching over the entire traffic from primary path to another backup path, the traffic through the faulty link is restored via another path, joining the two endpoints of the link \cite{link}. The other links on that primary path remain as it is. In this strategy, every link is provisioned with a backup path. The link protection schemes are fast because of the faster fault localization. Two of the link protection schemes are Ring cover \cite{RC1} and p-cycles. In Ring Cover, cyclic paths are identified in the network such that each edge is traversed by at least one cycle. In this scheme, some of the links may be covered with more than one cycle, which results in additional redundancy/ required spare capacity. Therefore to minimize spare capacity, the double-cycle ring cover was proposed by Vinodkrishnan {\em et.al.} \cite{RC2}. Each edge is provided with protection by two unidirectional cycles, each one in the opposite direction on the common link. The spare capacity reduces to $2N$ in this configuration. There are no cycles formed with the working capacity as done in Ring Cover protection. One of the drawbacks of double cycle ring covers is different signal delays in forward and backward directions after restoration on link failure. Further, the restoration times are also different in the two directions.

In ring cover and double cycle ring cover, only on-cycle failures are restored. If the straddling link\footnote{only end nodes are the part of the cycle, and not the link} failures can also be restored by the cycle, then we can further reduce the spare capacity requirement. This kind of configuration is called p-Cycle based protection. The concept of p-cycles was first introduced by Grover {\em et. al.} \cite{p1}. p-Cycles in an optical mesh network provide the same switching speed as the ring based protection $(< 50ms)$\footnote{Bi-directional Line Switched Ring} and  capacity efficiency as in mesh networks \cite{p1}, \cite{p2}. Whenever there is a link failure, each p-cycles, based on whether the failed link is on-cycle or straddling to the p-cycle, will provide a single unit (on-cycle) or two units (straddling\footnote{A straddling link must have its edge nodes on the p-cycle, but it's not part of the cycle.}) of protection to the working capacity.

A good amount of work has been done in the past that highlights the advantages of p-cycle for the protection and restoration of traffic \cite{p1}, \cite{p2}. p-Cycles can also be used for protecting nodes \cite{node}, paths \cite{FIPP, FIPP1} and path-segments \cite{ps} in addition to the links in an optical mesh network. For $100\%$ single link protection, optimum required p-cycles can be computed using an ILP (integer linear program). The main objective of optimization of the number of the p-cycles is to provide full protection while minimizing the required spare capacity.

The concept of the Hamiltonian cycle\footnote{cycle that traverses each node once} was introduced in the literature  by Sack {\em et. al.} \cite{h1} for a single link failure scenario when all the links are carrying same traffic to be protected. It is an efficient solution when all the links are almost equally loaded, as compared to the protection using multiple simple arbitrary p-cycles-based protection. It also mitigates the problem of loop-back as shown by Asthana {\em et. al.} \cite{h2} in her work.

The p-cycle-based protection in large networks has also been explored using the Multi-Domain approach \cite{MD1}, \cite{MD2}, \cite{MD3}, \cite{MD4}, \cite{MD5}. The idea is to partition the whole network into multiple domains using graph clustering algorithms. p-Cycles can be computed in each domain independently, forming intra-domain p-cycles. For protecting the links connecting different domains, separate inter-domain p-cycles are computed. Normally, each domain is an independently administered entity, and hence they are clearly defined by using some parameters like hop length. The time taken by individual (small-sized) domains to calculate the number of p-cycles (or spare capacity) is also expected to be less.

\section{Advance Technologies to Support Higher Data Rates in Future Optical Networks}
  
  \subsection{Distance-Adaptive Modulation Formats}
    
    \begin{figure}[ht!]
    \centering
        \includegraphics[width=0.8\linewidth]{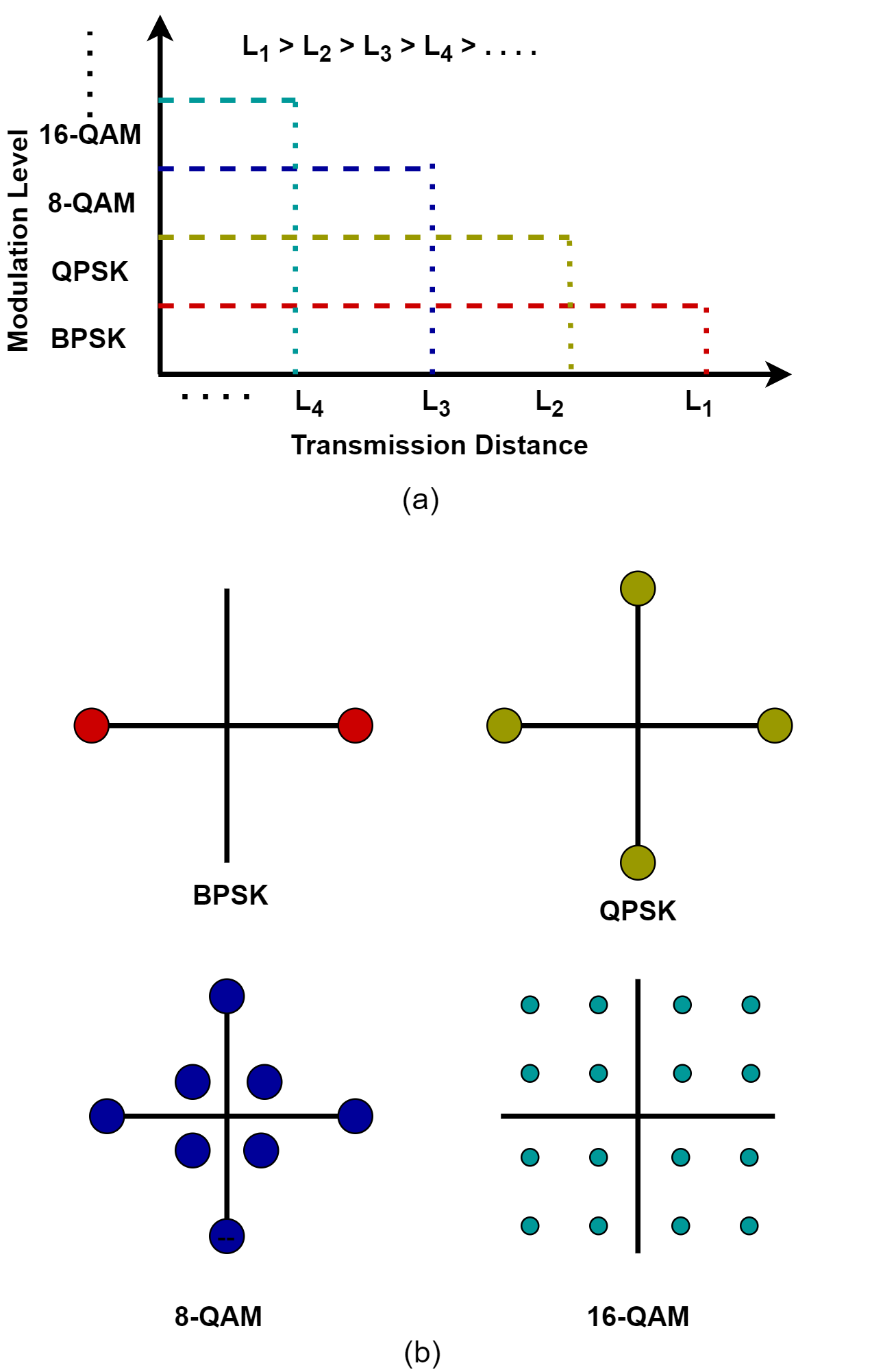}
        \caption{Constellation diagrams of modulation formats and their transmission reach.}
        \label{fig:AMF}
    \end{figure}
    
Using coherent detection techniques, a receiver can detect the amplitude as well as the phase of the received signal \cite{Coh}. Advanced modulation techniques which use both amplitude and phase modulation can carry more than one bit per symbol. The constellation of four advanced modulation formats is illustrated in Fig. \ref{fig:AMF}(a). Here the first number/letter represents the number of quantization levels ($N$). For a binary transmission system, the number of bits per symbol is calculated as $M$ = $log_2N$. Larger the value of $N$, the closer the symbols are placed in the constellation diagram. Signal in optical fiber deteriorates due to nonlinear impairments like self-phase modulation (SPM) and cross-phase modulation (XPM). Further, the amplitude of the received signal gets deteriorated due to Four-Wave-Mixing (FWM) and Amplified Spontaneous Emission (ASE) noise. Both the phase noise and amplitude noise accumulates over the transmission length. The higher the modulation format, the higher will be its Signal to Noise Ratio requirement to maintain the minimum Bit Error Rate required at the receiver side \cite{Impairment}. Therefore, the transmission reach of a modulation format reduces as we use a higher value of M, as shown in Fig. \ref{fig:AMF}(b). 

By using higher order modulation formats for shorter paths, the transmission rate of the single slot of 12.5 GHz can be increased manifold ($12.5 \times M$), as shown in Table \ref{tab:modulation}.  Hence, for shorter reach links the $SS_r$ requirement reduces by a factor of $M$, according to the equation,
\begin{equation}
  SS_r=\left\lceil\frac{b}{12.5 \times 2 \times M}\right\rceil,  
\end{equation}
where $b$ is the data rate required by a connection request. Accordingly, the average bandwidth requirement of the connection requests in the network reduces significantly, which in turn increases the spectrum efficiency. As a result, a lot of spectrum is saved to serve future connection requests.
    

Further, by using dual-polarization multiplexing (DPM), same spectrum slot can be used to transmit two individual signals in orthogonal polarizations. Using DPM, the spectral efficiency in bits/sec/Hz gets doubled. The transmission rate supported by various modulation formats using DPM against the maximum transmission reach is given in Table \ref{tab:modulation} \cite{AMF}.

\begin{table}[ht]
\caption{Modulation formats against path length}
\begin{center}
\begin{tabular}{|c|c|c|c|}
\hline
Modulation & Bits per & Data & Maximum\\
Format & Symbol (M) & Rate (Gbps) &Reach (Km)\\
 \hline
DP-BPSK & 1 & 25 & 8000 \\
\hline
DP-QPSK & 2& 50 & 4000\\
\hline
DP-8QAM & 3 & 75 & 2000\\
\hline
DP-16QAM & 4 & 100 & 1000\\
\hline
DP-32QAM & 5 & 125 & 500\\
\hline
DP-64QAM & 6 & 150 & 250\\
\hline
\end{tabular}
\label{tab:modulation}
\end{center}
\end{table}

The resource allocation algorithm for elastic optical networks incorporating distance adaptive modulation formats is called, Routing, Modulation, and Spectrum Assignment (RMSA) \cite{RMSA}. RMSA algorithms exhibit higher spectrum efficiency than both RWA and RSA algorithms discussed in section III.
    
    \subsection{Migration to Space Division Multiplexed-Optical Networks}
   The transmission capacity of a single-core fiber will soon approach its fundamental limits due to an increase in bandwidth demand by services and degradation of the transmission signal by nonlinear optical effects\cite{nonlinear}. One of the possible solutions which has been extensively researched in the last decade is the introduction of multiple spatial channels, which can increase the capacity of the optical networks. The idea of expansion of the transmission dimension from single core/mode to multiple cores/modes is called Space Division Multiplexing (SDM). Integration of Space Division Multiplexing with a suitable optical network transmission technology can help the network to match the pace of internet traffic growth by increasing the transmission capacity. Different types of fibers shown in Fig. 24, i.e., Single-Mode fiber bundle, Multi-core fiber, and Few-mode multi-core fiber, are used in SDM technology\cite{sdm}.\\
   \begin{figure}[ht!]
    \centering
        \includegraphics[width=0.9\linewidth]{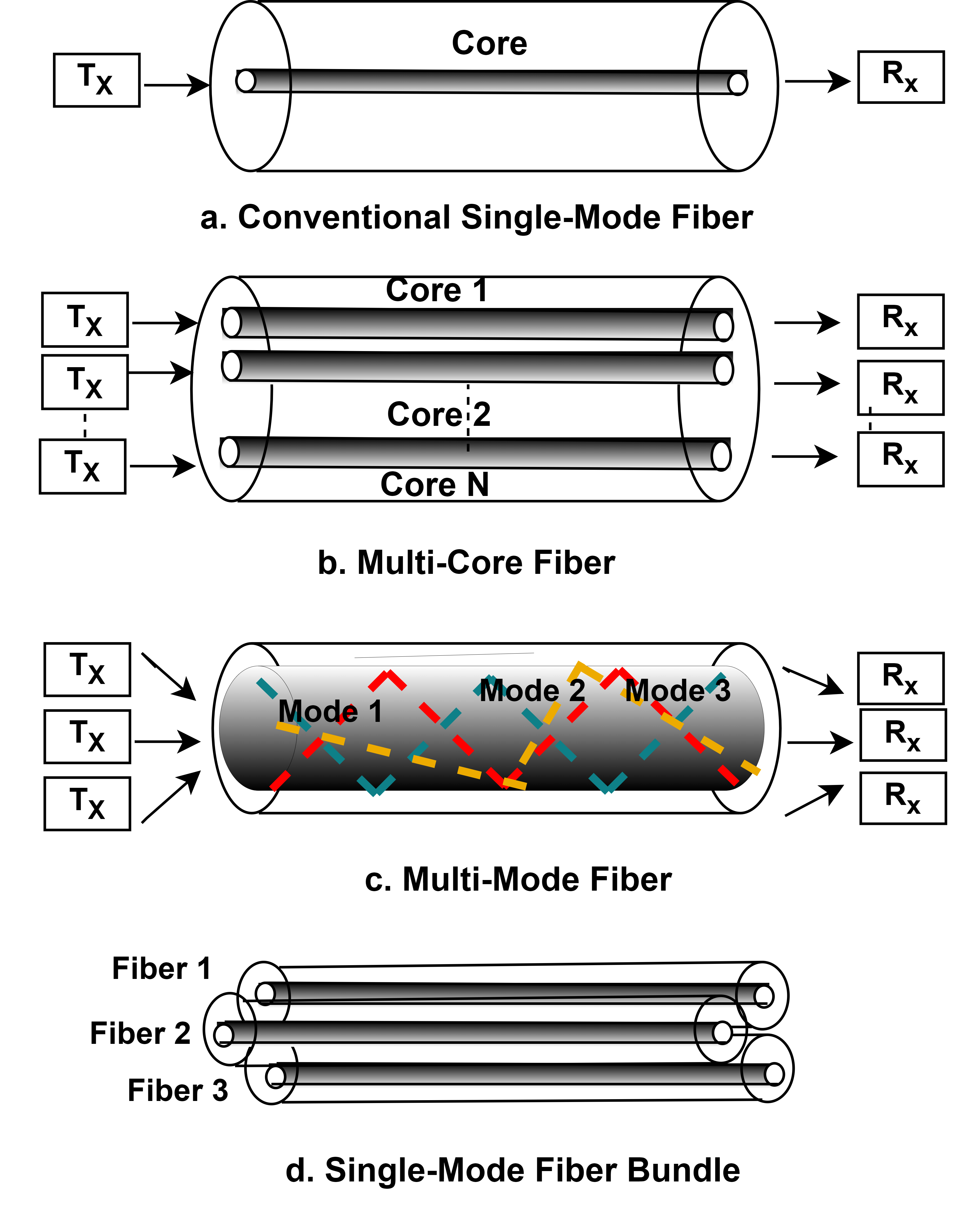}
        \caption{Different types of fibers used in SDM technology}
        \label{fig:MCF}
    \end{figure}
    \begin{enumerate}  
    
   \item Single-Mode Fiber Bundle (SMFB): An optical fiber cable generally consists of a bundle of multiple fibers packed together. Depending on the application, the cable may consist of ten to hundred fibers. Fibers not currently being used for transmission are called "black fibers". Since installing a new fiber cable is very costly and time-consuming, these extra fibers are already laid in the existing core optical network for future usage. Therefore, the easiest option to increase the capacity of an optical transport link is to use these black fibers. Another advantage of using multiple fiber bundles is using readily available transmitters, receivers, amplifiers to set up a new transmission link. A single-core $\&$ single-mode fiber can be easily spliced together in case of a fiber-cut. There are some disadvantages too. Using multiple independent fibers will also require as many numbers of network components like transceivers and amplifiers. As a result, the cost/per bit will increase linearly. Users may not be willing to pay high cost/bit for future high data rate applications. Also, the size of the cable will become very thick if a large number of fibers are placed inside it. It may be difficult to handle the cable and to install the cable in space-limited areas. Therefore, SMFB-based SDM technology seems lucrative for near-future goals. But for cost and energy-efficient future SDM optical networks, SMFB technology may not be scalable. SMFB-based Optical networks using both SDM and WDM are known as SDM-WDM optical networks \cite{SWDM1, SWDM2}. The resource allocation algorithm for SDM-WDM optical networks is called Routing, Core, and Wavelength Assignment (RCWA) \cite{RCWA}.    

   \item Few-Mode Fiber (FMF): A FMF can also be used to implement spatial multiplexing. The FWFs used for SDM are designed to support a few degenerated mode groups. These mode groups travel with different propagation constants inside the fiber core. These modes are multiplexed together to transmit independent channels. Since these modes couple with each other while propagating, they can only be routed to a single destination, where MIMO techniques are used to decouple and extract independent data streams. Due to the co-propagation of many modes inside the single core, FMFs suffer from impairments such as differential group delay,  differential attenuation, and modal interference, making them more suitable only for short-reach transmission \cite{FMF}.

   \item Multi-Core Fiber (MCF): MCF consists of Multiple single-mode cores fabricated inside a single cladding \cite{19core}. Cores placed inside a fiber can be coupled or uncoupled based on the distance of the centre of one core from another. Ultra-low crosstalk and low-loss MCF have been designed and fabricated\cite{MCF}, which seems to be one of the most popular ways to realize SDM. MCFs can provide very high capacity, up to Petabits per second, and have better physical dimension efficiency. 7-cores\cite{MCF} MCFs with hexagonal close-packed structures, shown in Fig. 25, have been fabricated and used in several experiments.
   
   \item Few-Mode Multi-Core Fiber (FM-MCF): A FM-MCF has multiple cores, and in each core, multiple independent channels can be transmitted in different mode groups using mode-division multiplexing. The impairments due to coupling among multiple modes can be compensated with Multi-input Multi-output digital signal processing (MIMO-DSP) on the receiver side\cite{MIMO-DSP}. To relax the MIMO-DSP requirement, authors in \cite{FM-MCF} have designed and fabricated weakly-coupled FM-MCF to suppress inter-mode coupling. Thus, the capacity of the FM-MCF fibre increases proportionally to the number of cores and and number of modes inside a core.
   \end{enumerate}

   Among the different SDM enabling techniques discussed above, Multi-core fiber-based SDM is researched primarily due to being more space and energy efficient. Multi-Core Optical Fibers have approximately the same attenuation profile as conventional single-mode fibers. Unlike traditional single-core fiber, the cores of MCFs work as independent fiber with their own set of spectrum slots. MCFs are fabricated using stack-and-draw technology\cite{stackanddraw}, which involves three significant steps, fabrication of a single-mode fiber preform, stacking of multi-core preform, and fiber drawing of MCFs. Several MCFs have been designed are proposed in the literature for SDM application, and they can be categorized based on various fiber attributes and parameters like refractive index profile (MCFs with step-index and graded-index), core pitch (MCFs with coupled-core and uncoupled-core), spatial homogeneity (MCFs with homogeneous and heterogeneous cores), and modal regime (MCFs with single-mode and multi-mode).
   
   Some challenges in the designing of MCFs, as described below,

    \begin{enumerate}[label=\Roman*.]       \item Fabrication of MCFs with a maximum number of cores in the same cladding diameter without deteriorating the transmission signal quality,
   
        \item To provide maximum core isolation for minimizing the Inter-Core Crosstalk(IC-XT) impairment (discussed in detail in the following subsection).
   \end{enumerate}
   
   \begin{figure}[ht!]
   \centering
\includegraphics[width=0.5\linewidth]{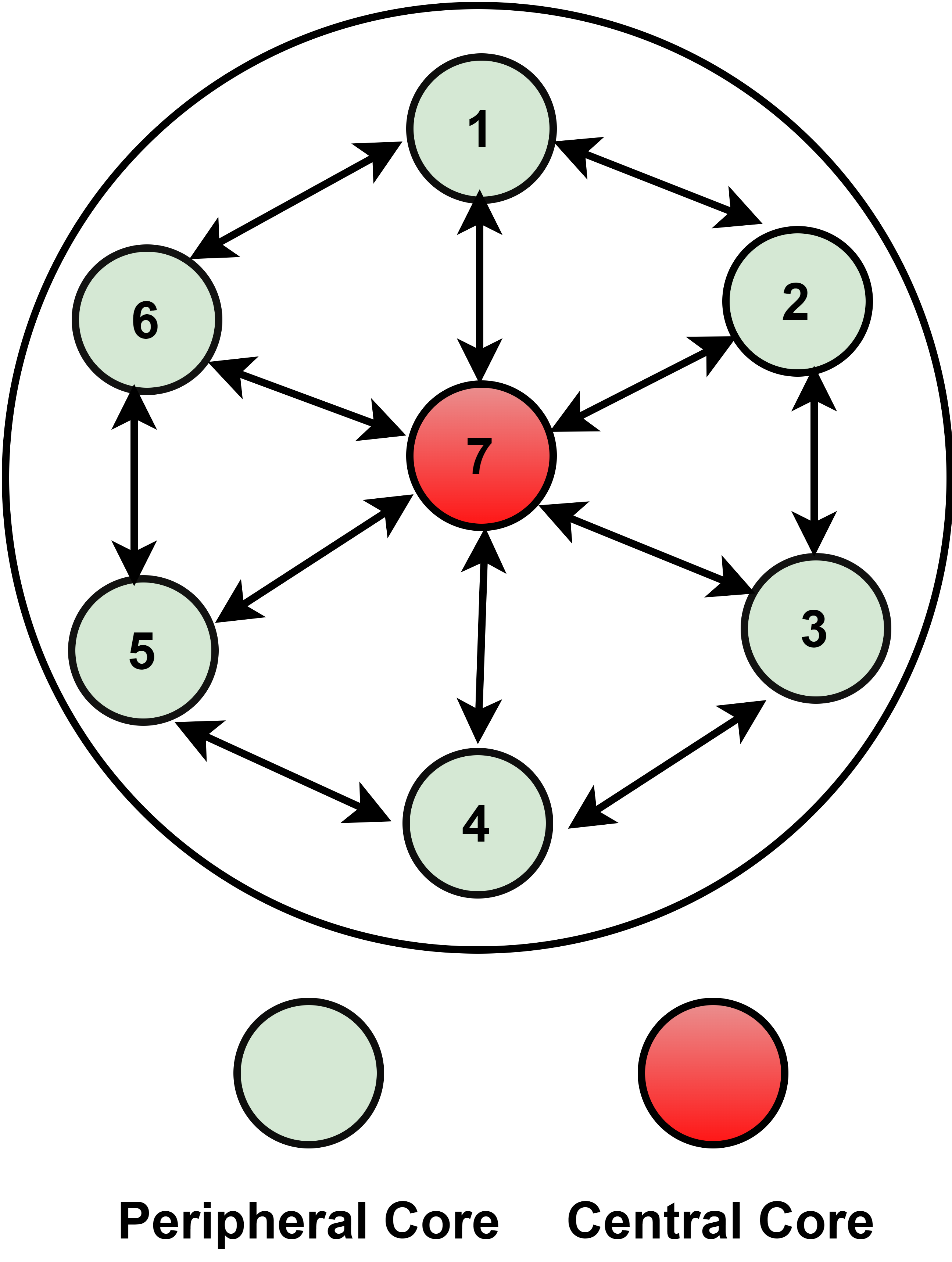}
\caption{A 7-Core Multi-Core Fiber}
\label{fig:MCF}
\end{figure}
    
\subsubsection{Crosstalk in MCF-based SDM}
 In MCFs, power leakage is observed among the adjacent cores inside the fiber. The major portion of the electromagnetic field corresponding to a guided mode remains inside the fiber core, and some fraction of it overlaps into the adjacent cores. The field inside the core is known as the transmission optical field, and the field in the cladding is known as the evanescent field and leads to leakage loss. The Evanescent field coupling between cores is called Inter-Core Crosstalk (IC-XT). If IC-XT is high, it may severely degrade the signal quality at the other end of fiber. Quantitatively IC-XT is evaluated using the following equation, 
\begin{equation}
XT=\frac{n-n.exp[-(n+1).2hL]}{1+n.exp[-(n+1).2hL]},   
\end{equation}
where
\begin{equation}
    h=\frac{2k^2r}{\beta\Lambda}.
\end{equation}

Here, $k$, $\beta$, $r$, and $\Lambda$ are the fiber coupling coefficient, propagation constant, fiber bending radius, and the core pitch (distance between the center of the two cores). $n$ is the number of adjacent cores to the core under consideration, and $L$ is the length of the fiber. For a 7-Core MCF, as shown in Fig. 25, the IC-XT in the central core is due to 6 adjacent peripheral cores, whereas in the peripheral cores, it is due to 3 adjacent cores.\\

\begin{figure}[ht!]
    \centering
\includegraphics[width=0.8\linewidth]{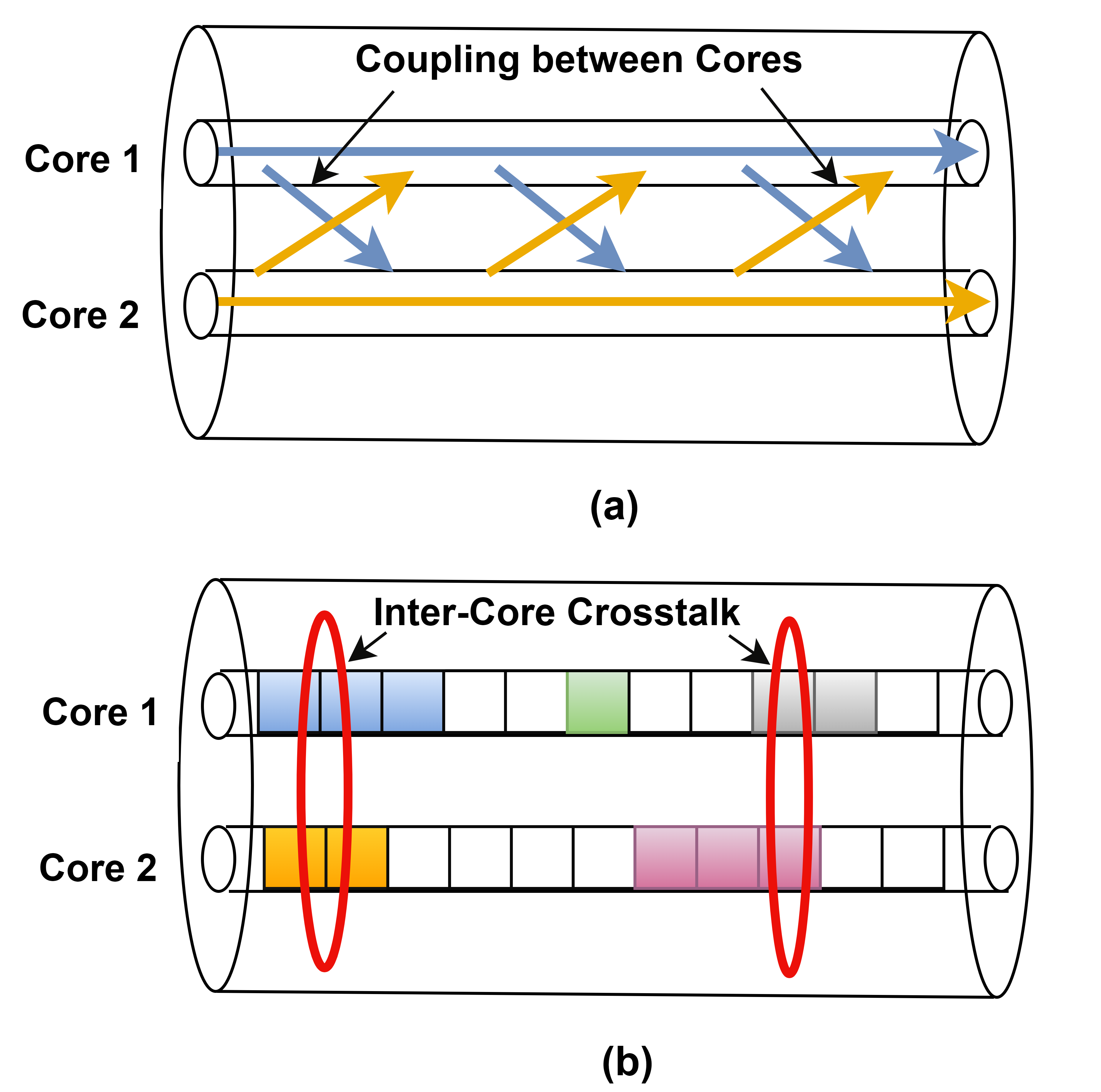}
        \caption{(a) Coupling between cores in Multi-Core Fiber
        (b) Illustration of Inter-Core Crosstalk in Multi-Core Fibers}
        \label{fig:MCF}
    \end{figure}
    
 is more intense when spectrum slots are occupied in the adjacent cores have same-index, as shown in Fig. 26(b). IC-XT is negligible when same-indexed spectrum slots are occupied in the non-adjacent cores. No IC-XT exists when occupied spectrum slots are at different indices in adjacent cores. Significant research has been done about different methods to mitigate IC-XT. Some of them are\\

\begin{itemize}
    \item Trench-assisted Multi-core fiber.
    \item Hole-assisted Multi-core fiber.
    \item Design of an appropriate Routing, Modulation, Core, and Spectrum Assignment (RMCSA) techniques for assigning resources in the most optimal manner considering crosstalk.
    \item MCFs with bidirectional assignment between neighbouring cores.
\end{itemize}


Trench-assisted Multi-core fiber (Fig. 27) has been demonstrated to reduce crosstalk compared to step-indexed MCFs \cite{MCF}. The overlapping evanescent field between adjacent cores is reduced due to the suppression of the transmission field distribution in each core owing to the trench index. Hole-assisted structures around each core, also known as multi-core hole-assisted fibers (MC-HAFs) shown in Fig.27, have also been shown  to suppress crosstalk between neighbouring cores compared to conventional MCFs \cite{HAMCF}.\\ 
\begin{figure}
\centering
\includegraphics[width=0.9\linewidth]{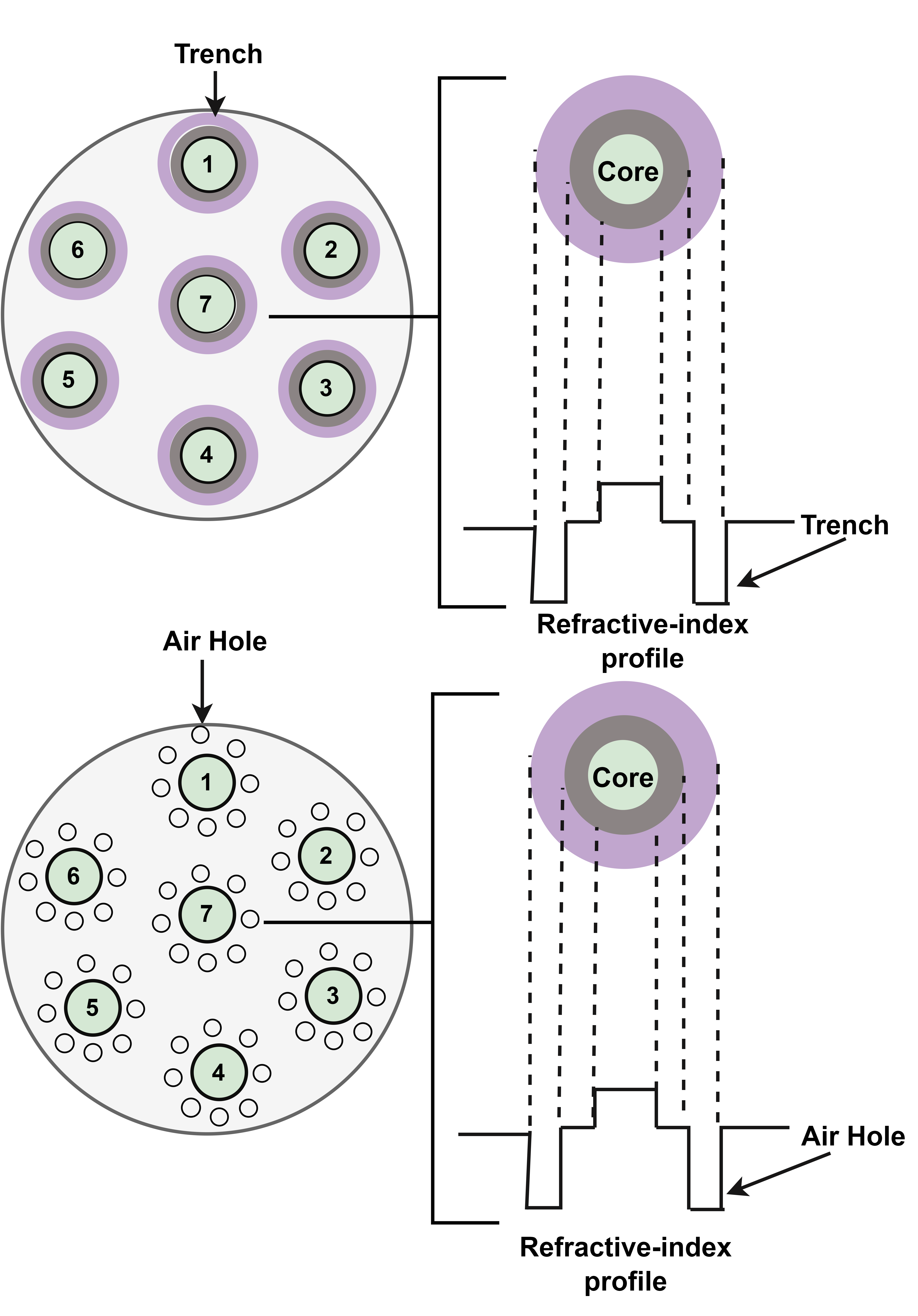}
\caption{Trench and Hole assisted Multi-Core Fiber}
\label{fig:MCF}
\end{figure}


Whenever a request with some bandwidth demand arrives at a node, the network tries to establish a lightpath between the source and destination, i.e., Routing. According to the path length, it selects the most suitable modulation format to determine the required number of spectral slots as per the following equation,
\begin{equation}
SS_r=\left\lceil\frac{b}{12.5*M}\right\rceil.  
\end{equation}
If we consider polarization multiplexing, the above equation gets modified as follows
\begin{equation}
  SS_r = \left\lceil\frac{b}{12.5 \times 2 \times M}\right\rceil.  
\end{equation}

Here $SS_r$, $b$, and $M$ are the required number of spectral slots, bandwidth demand per request, and modulation format (m = 1, 2, 3, 4, respectively for BPSK, QPSK, 8-QAM, and 16-QAM). Since cores add a new spatial dimension, we need to search the cores for resource assignment after the best path/route and the required number of slots are determined.\\

\begin{figure}[ht!]
\centering
\includegraphics[width=1\linewidth]{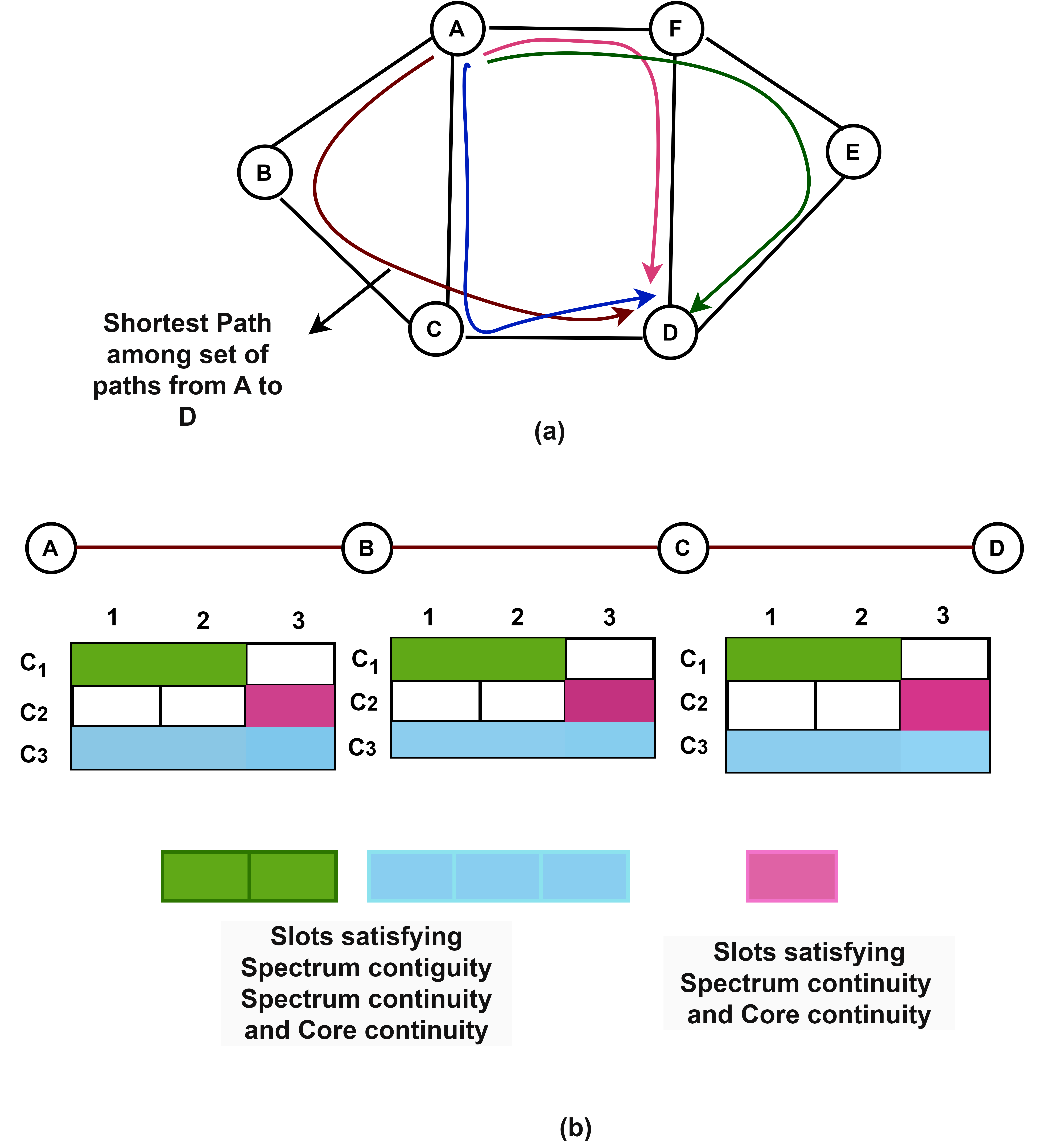}
\caption{(a) Route finding for connection requests from source ``A" to ``D" (b) Resource assignment satisfying all constraints. }
\label{fig:Route}
\end{figure}

Fig. \ref{fig:Route} shows an optical network topology with six nodes and eight bi-directional multi-core fiber links. Each multi-core fiber has three cores ($c_1$, $c_2$, and $c_3$), each with three spectral slots (assumed for simplicity). Three connection requests with required spectral slots (2, 1, and 3) are allocated in the shortest path (Demand from A to D), satisfying all the constraints,i.e., Spectrum contiguity, Spectrum continuity, and Core continuity. Core continuity ensures that the same core is used in all the links spanning the lightpath. The RMCSA algorithms should also check whether or not the crosstalk conditions are met for the chosen route. Using appropriate crosstalk-aware RMCSA strategies can suppress the impairment due to IC-XT. In \cite{XT-AWARE} and \cite{COMSNETS_b}, work is done to allocate resources in cores based on predefined core priorities. This is a best-effort avoidance approach, where efforts are made to allocate the resources in non-adjacent cores on a priority basis. Crosstalk-aware RMCSA algorithms \cite{crosstalk},\cite{rmcsa} also reduce the effect of the crosstalk compared to the allocation of the resources using first-fit strategy. Various Machine Learning-aided resource assignments have been proposed in SDM-EONs. A Spectrum partitioning scheme for adjacent cores of an MCF, proposed in \cite{ML1}, has been shown to improve spectrum utilization while reducing inter-core crosstalk. A Crosstalk-aware RMCSA algorithm known as Tridental Resource Allocation (TRA)\cite{ml2} has also been proposed, which maintains a good balance between spectrum utilization and crosstalk in the network. Furthermore, in \cite{ml3}, TRA and its variant translucency-aware TRA (TaTRA) algorithm have been proposed for transparent and translucent Spectrally Spatially Flexible Optical Networks. Authors in \cite{DELCON_b} have proposed a fragmentation-aware and crosstalk-ignore-based RCSA algorithm for SDM-EON networks based on uncoupled-core MCFs. The above RMCSA algorithms have been shown to perform better than existing algorithms in the literature. 

\begin{figure}[ht!]
\centering
\includegraphics[width=0.9\linewidth]{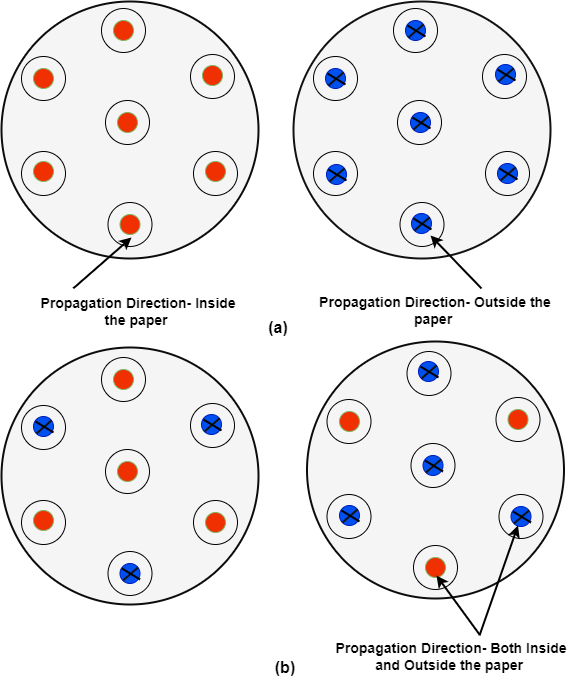}
\caption{(a) Co-directional propagation in a Multi-Core fiber (b) Contra-directional propagation in a Multi-Core fiber }
\label{fig:MCF}
\end{figure}
The concept of bidirectional propagation in optical fibers has also been researched in the context of MCFs (Fig. \ref{fig:MCF}). A pair of MCFs carries traffic in both directions between nodes. But MCFs with bidirectional assignment\cite{bidirec1},\cite{bidirec2},\cite{bidirec3}, i.e., cores inside the same fiber carry traffic in different directions, have been shown to reduce the crosstalk compared to unidirectional traffic assignment. The reason behind this is that in contra-directional propagation, optical signals travel in the opposite direction, and for a very short interval of time, crossover occurs between them; hence, the crosstalk built is minimal compared to co-directional propagation.

\section{Multi-Band Optical Networks}

Among the solutions proposed in the literature to solve the capacity crunch problem, one of the solutions extensively researched in the past decade is multiband \cite{mb1} technology in the optical network. In MB, the idea is to exploit the existing capacity in the optical fiber beyond C-band (1530–1565 nm) without deploying additional fibers.

Earlier, only the L-band was used in conjunction with the C-band. The advantage of using the L-band (1565–1625 nm) is that it uses an Erbium-Doped Fiber Amplifier \cite{mb2}, which has been used in the C-band \cite{mb3}. 

To expand the capacity further, the researchers have been measuring the performance of other bands with C+L bands, for example, S-band (1460–1530 nm) and E-band (1360–1460 nm). N. Sambo et al. \cite{mb4} compared the CLS and CLE-based networks. The various network performances have been evaluated to determine which additional band provides better Quality of Transmission \cite{mb4}. Their analysis has shown that CLE with a Guard Band (GB) of 14 THz between C+L and E performs better than CLS and CLE (with lower GB). Selecting a band is also critical from an architectural point of view since for the S-band, a Thulium-Doped Fiber Amplifiers (TDFA) \cite{mb5}, \cite{mb6}, and the E-band, Bismuth-Doped Fiber Amplifiers (BDFA) \cite{mb7} and Nd3+ Doped Fiber Amplifiers (NDFA) \cite{mb8} are needed. 

Similarly, the use of the O-band (1260–1360 nm) with the C + L band is explained by V. Mikhailov et al., where a Bismuth Doped Fiber Amplifier is used \cite{mb9} for amplification. 
 
Using multiband in Optical Networks came with challenges, such as multiband node architecture, inter-band impairments, and routing and resource allocation. There are research works that focus on these problems. GSNR is used to check the performance of the multiband transmission. It takes into account both linear and Non-Linear effects \cite{mb10}.

\begin{figure}[ht!]
\centering
\includegraphics[width=1\linewidth]{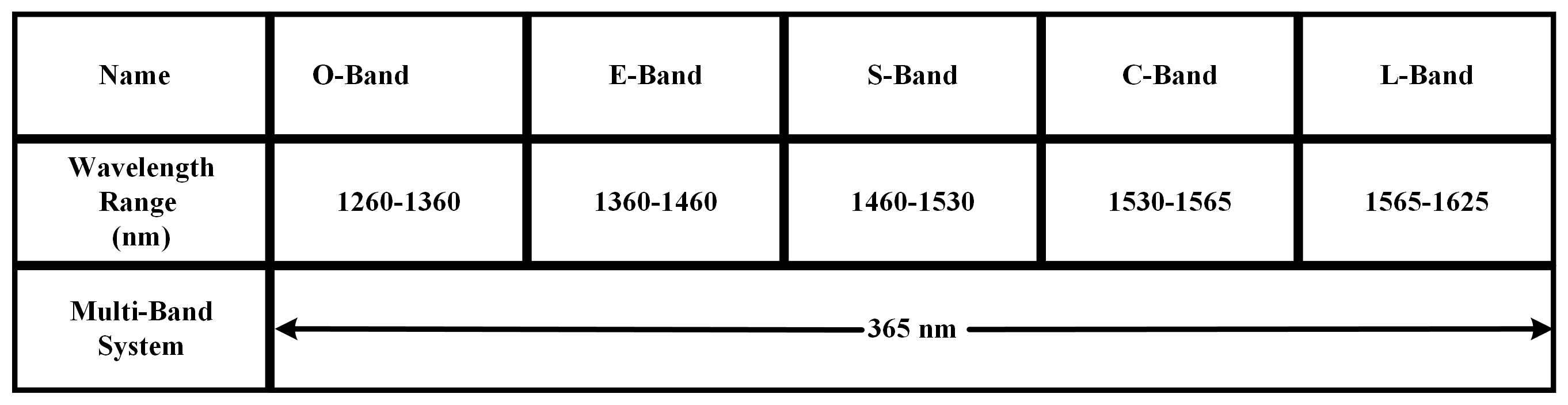}
\caption{}
\label{fig:MB}
\end{figure}

\section{Future Prospective}
\begin{itemize}
\item Designing of optimized Routing and Resource Assignment Algorithms considering spectral and spatial dimensions for high capacity transmissions.
\item Node architecture for such a high capacity nodes (increase in the number of ports) and solution to reduce load of Wavelength Selective Switches (WSSs). 
\item Beside protection, faster restoration mechanisms fir such an optical networks.

\end{itemize}
\section{Concluding Remarks}
The Optical Networks form the core of the communication networks. Over the years, many technological advancements have made it an unbeatable choice for the transmission of aggregated data.

\end{document}